\documentclass[longauth]{aa}

\usepackage[T1]{fontenc}
\usepackage{ae,aecompl}
\usepackage{natbib,twoopt}
\usepackage{booktabs}
\usepackage{ulem}
\usepackage{placeins}
\usepackage{caption}
\usepackage{booktabs}

\usepackage{graphicx}   
\usepackage{amsmath}    
\usepackage{amssymb}    
\usepackage{gensymb}    
\usepackage{multirow}
\usepackage{textcomp} 
\usepackage{mathtools}
\usepackage{IEEEtrantools}
\usepackage{float}
\usepackage{rotating}
\usepackage{bm}
\usepackage{adjustbox}             
\usepackage{subcaption}            
\usepackage{threeparttable}
\usepackage{txfonts} 
\usepackage{footmisc}
\usepackage{hyperref}
\hypersetup{
    colorlinks=true,
    linkcolor=blue,
    filecolor=blue,
    urlcolor=blue,
    citecolor=blue,
}
\usepackage{comment}
\usepackage{pdflscape}
\usepackage{placeins}
\usepackage{scalefnt}
\usepackage{xcolor}                 

\usepackage{relsize}

\newcommand{\Msun}{\hbox{$\mathrm{M}_{\rm{\odot}}$}}
\newcommand{\angstrom}{\text{\normalfont\AA}}

\newcommand{\tgrb}{\hbox{$T_{90}$}}

\newcommand{\halpha}{\hbox{${\mathrm{H}}{\alpha}$}\xspace}
\newcommand{\hbeta}{\hbox{${\mathrm{H}}{\beta}$}\xspace}
\newcommand{\paalpha}{\hbox{${\mathrm{Pa}}{\alpha}$}\xspace}
\newcommand{\OII}{\hbox{[O\,\textsc{ii}]\,$\lambda3727$}\xspace}
\newcommand{\OIIIa}{\hbox{[O\,\textsc{iii}]\,$\lambda4959$}\xspace}
\newcommand{\OIIIb}{\hbox{[O\,\textsc{iii}]\,$\lambda5007$}\xspace}
\newcommand{\OIIIab}{\hbox{[O\,\textsc{iii}]\,$\lambda\lambda4959,5007$}\xspace}
\newcommand{\OIII}{\hbox{[O\,\textsc{iii}]}\xspace}
\newcommand{\NIIb}{\hbox{[N\,\textsc{ii}]\,$\lambda6583$}\xspace}

\newcommand{\SVOM}{SVOM\xspace}
\newcommand{\Swift}{\textit{Swift}\xspace}

\newcommand{\EinsteinProbe}{\textit{Einstein Probe}\xspace}
\newcommand{\EP}{\textit{EP}\xspace}
\newcommand{\JWST}{\textit{JWST}\xspace}

\newcommand{\orcid}[1]{\href{https://orcid.org/#1}{\includegraphics[width=8pt]{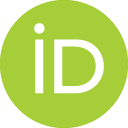}}}

\graphicspath{{Figures/}} 

\begin{document}
    \title{XRF~241001A/SN~2024aiiq: A faint soft X-ray transient detected by SVOM with a broad-line type Ic supernova revealed by JWST}

\author{
B.~Schneider\orcid{0000-0003-4876-7756}\inst{1}\fnmsep\thanks{E-mail: benjamin.schneider@lam.fr},
M.~Brunet\orcid{0009-0002-6036-9873}\inst{2},
B.~P.~Gompertz\orcid{0000-0002-5826-0548}\inst{3,4},
D.~Turpin\orcid{0000-0003-1835-1522}\inst{5},
D.~B.~Malesani\orcid{0000-0002-7517-326X}\inst{6,7,8},
O.~Godet\orcid{0000-0001-7635-9544}\inst{2},
A.~J.~Levan\orcid{0000-0001-7821-9369}\inst{8,9},
F.~Daigne\orcid{0009-0005-7119-4716}\inst{10,11},
N.~Sarin\orcid{0000-0003-2700-1030}\inst{12,13},
N.~A.~Rakotondrainibe\orcid{0009-0004-0263-7766}\inst{1},
A.~Martin-Carrillo\orcid{0000-0001-5108-0627}\inst{14},
J.~T.~Palmerio\orcid{0000-0002-9408-1563}\inst{5},
C.~C.~Th\"one\orcid{0000-0002-7978-7648}\inst{15},
H.~L.~Li\orcid{0000-0005-7119-4716}\inst{16},
A.~Saccardi\orcid{0000-0002-6950-4587}\inst{5,17},
A.~de~Ugarte~Postigo\orcid{0000-0001-7717-5085}\inst{1},
S.~Antier\orcid{0000-0002-7686-3334}\inst{18},
V.~Buat\inst{1},
D.~{\v D}urov{\v c}{\'i}kov{\'a}\orcid{0000-0001-8986-5235}\inst{19,20},
L.~Izzo\orcid{0000-0001-9695-8472}\inst{21,22},
J.~K.~Leung\orcid{0000-0002-9415-3766}\inst{23,24,25},
G.~Mo\orcid{0000-0001-6331-112X}\inst{26,27},
Y.~L.~Qiu\inst{16},
S.~D.~Vergani\orcid{0000-0001-9398-4907}\inst{28},
J.~Wang\inst{16,29},
J.~Y.~Wei\inst{16,29},
L.~P.~Xin\orcid{0000-0002-9422-3437}\inst{16,29},
R.~Mochkovitch\inst{10},
B.~Zhang\orcid{0000-0002-9725-2524}\inst{30,31,32,33},
H.~B.~Cai\inst{16},
S.~Campana\orcid{0000-0001-6278-1576}\inst{34},
A.~Coleiro\orcid{0000-0003-0860-440X}\inst{35},
B.~Cordier\inst{5},
P.~D’Avanzo\inst{36},
N.~Dagoneau\orcid{0000-0002-1361-2562}\inst{37},
V.~D'Elia\inst{38},
Y.~W.~Dong\inst{39},
D.~G\"otz\orcid{0000-0001-9494-0981}\inst{5},
S.~Guillot\orcid{0000-0002-6449-106X}\inst{2},
X.~H.~Han\orcid{0000-0002-6107-0147}\inst{16,30},
D.~H.~Hartmann\orcid{0000-0002-8028-0991}\inst{40},
L.~Huang\orcid{0000-0003-3882-8316}\inst{16},
Y.~F.~Huang\orcid{0000-0001-7199-2906}\inst{41,42},
P.~Jakobsson\orcid{0000-0002-9404-5650}\inst{43},
A.~Klotz\orcid{0000-0003-0106-4148}\inst{2},
C.~Lachaud\orcid{0000-0002-3732-854X}\inst{36},
E.~Le~Floc'h\orcid{0000-0001-7421-4413}\inst{5},
X.~M.~Lu\inst{16},
P.~Maggi\orcid{0000-0001-5612-5185}\inst{44},
M.~De~Pasquale\inst{45},
F.~Piron\orcid{0000-0001-6885-7156}\inst{46},
R.~Salvaterra\orcid{0000-0002-9393-8078}\inst{47},
S.~Schanne\orcid{0009-0007-1271-9900}\inst{5},
J.~Sollerman\orcid{0000-0003-1546-6615}\inst{48},
N.~R.~Tanvir\orcid{0000-0003-3274-6336}\inst{49},
Z.~Vidadi\inst{50},
P.~Wang\orcid{0000-0003-0466-2223}\inst{39},
C.~Wu\orcid{0009-0001-7024-3863}\inst{16,30},
S.~L.~Xiong\inst{39},
Y.~Xu\inst{16},
T.~Zafar\orcid{0000-0003-3935-7018}\inst{51},
P.~P.~Zhang\inst{16},
S.~N.~Zhang\orcid{0000-0001-5586-1017}\inst{39,30},
S.~J.~Zheng\orcid{0000-0003-2256-6286}\inst{39}
}

\institute{
Aix Marseille University, CNRS, CNES, LAM, Marseille, France
\and 
IRAP, Université de Toulouse/CNRS/CNES, 9 avenue du colonel Roche, 31028 Toulouse, France
\and 
School of Physics and Astronomy, University of Birmingham, Birmingham, B15 2TT, UK
\and 
Institute for Gravitational Wave Astronomy, University of Birmingham, Birmingham, B15 2TT, UK
\and 
CEA Paris-Saclay, IRFU/DAp-AIM, 91191 Gif-sur-Yvette, France
\and 
Cosmic Dawn Center (DAWN), Denmark
\and 
Niels Bohr Institute, University of Copenhagen, Jagtvej 155A, Copenhagen N, DK-2200, Denmark
\and 
Department of Astrophysics/IMAPP, Radboud University, PO Box 9010, 6500 GL Nijmegen, The Netherlands
\and 
Department of Physics, University of Warwick, Gibbet Hill Road, CV4 7AL Coventry, United Kingdom
\and 
Sorbonne Université, CNRS, UMR 7095, Institut d'Astrophysique de Paris, 98 Bis bd Arago, 75014, Paris, France
\and 
Institut Universitaire de France
\and 
Kavli Institute for Cosmology, University of Cambridge, Madingley Road, CB3 0HA, UK
\and 
Institute of Astronomy, University of Cambridge, Madingley Road, CB3 0HA, UK
\and 
School of Physics and Centre for Space Research, University College Dublin, Belfield, Dublin 4, Ireland
\and 
E. Kharadze Georgian National Astrophysical Observatory, Mt. Kanobili, Abastumani 0301, Adigeni, Georgia
\and 
CAS Key Laboratory of Space Astronomy and Technology, National Astronomical Observatories, Chinese Academy of Sciences, Beijing 100101, China
\and 
Centre national d’études spatiales (CNES), Paris, France
\and 
IJCLab, Univ Paris-Saclay, CNRS/IN2P3, Orsay, France
\and 
MIT Kavli Institute for Astrophysics and Space Research, 77 Massachusetts Avenue, MA 02139, USA
\and 
Department of Physics, Massachusetts Institute of Technology, 77 Massachusetts Avenue Cambridge, MA 02139, USA
\and 
INAF, Osservatorio Astronomico di Capodimonte, Salita Moiariello 16, I-80121 Naples, Italy
\and 
DARK, Niels Bohr Institute, University of Copenhagen, Jagtvej 155A, 2200 Copenhagen, Denmark
\and 
David A. Dunlap Department of Astronomy and Astrophysics, University of Toronto, 50 St. George Street, Toronto, ON M5S 3H4, Canada
\and 
Dunlap Institute for Astronomy and Astrophysics, University of Toronto, 50 St. George Street, Toronto, ON M5S 3H4, Canada
\and 
Racah Institute of Physics, The Hebrew University of Jerusalem, Jerusalem 91904, Israel
\and 
Department of Astronomy, California Institute of Technology, Pasadena, CA 91125, USA
\and 
The Observatories of the Carnegie Institution for Science, Pasadena, CA 91101, USA
\and 
LUX, Observatoire de Paris, Université PSL, CNRS, Sorbonne Université, 92190 Meudon, France
\and 
School of Astronomy and Space Science, University of Chinese Academy of Sciences, Beijing 101408, China
\and 
The Hong Kong Institute for Astronomy and Astrophysics, The University of Hong Kong, Pokfulam, Hong Kong, China
\and 
Department of Physics, The University of Hong Kong, Pokfulam, Hong Kong, China
\and 
Nevada Center for Astrophysics, University of Nevada, Las Vegas, USA
\and 
Department of Physics and Astronomy, University of Nevada, Las Vegas, USA
\and 
INAF–Osservatorio Astronomico di Brera, Via E. Bianchi 46, 23807 Merate (LC), Italy
\and 
Universit\'e Paris Cit\'e, CNRS, Astroparticule et Cosmologie, F-75013 Paris, France
\and 
Como Lake centre for AstroPhysics (CLAP), DiSAT, Università dell’Insubria, via Valleggio 11, 22100 Como, Italy
\and 
CEA Paris-Saclay, IRFU/DEDIP, 91191 Gif-sur-Yvette, France
\and 
Space Science Data Center (SSDC) - Agenzia Spaziale Italiana (ASI), Via del Politecnico snc, I-00133 Roma, Italy
\and 
State Key Laboratory of Particle Astrophysics, Institute of High Energy Physics, Chinese Academy of Sciences, Beijing 100049, China
\and 
Clemson University, Department of Physics \& Astronony, Clemson, SC 29634, USA
\and 
School of Astronomy and Space Science, Nanjing University, Nanjing 210023, China
\and 
Key Laboratory of Modern Astronomy and Astrophysics (Nanjing University), Ministry of Education, China
\and 
Centre for Astrophysics and Cosmology, Science Institute, University of Iceland, Dunhagi 5, 107 Reykjavik, Iceland
\and 
Observatoire Astronomique de Strasbourg, Universit\'e de Strasbourg, CNRS, 11 rue de l’Universit\'e, F-67000 Strasbourg, France
\and 
University of Messina, Mathematics, Informatics, Physics and Earth Science Department, Via F.S. D’Alcontres 31, Polo Papardo, 98166, Messina, Italy
\and 
Laboratoire Univers et Particules de Montpellier, Universit\'e Montpellier, CNRS/IN2P3, F-34095 Montpellier, France
\and 
INAF-Istituto di Astrofisica Spaziale e Fisica Cosmica di Milano, Via A. Corti 12, 20133 Milano, Italy
\and 
Oskar Klein Centre, Department of Astronomy, Stockholm University, AlbaNova, SE-106 91 Stockholm, Sweden
\and 
School of Physics and Astronomy, University of Leicester, University Road, Leicester, LE1 7RH, UK
\and 
N. Tusi Shamakhy Astrophysical Observatory Azerbaijan National Academy of Sciences, settl.Mamedaliyev, AZ 5626, Shamakhy, Azerbaijan
\and 
School of Mathematical and Physical Sciences, Macquarie University, NSW 2109, Australia
}
   
    \date{Accepted XXX. Received YYY; in original form ZZZ}

  \abstract
   {X-ray flashes (XRFs) are a type of gamma-ray burst (GRB) with prompt emission predominantly below $\sim$30~keV and have been poorly detected by previous missions monitoring the high-energy sky. The advent of the Space-based multiband astronomical Variable Objects Monitor (\SVOM), with its wide-field instrument ECLAIRs, provides a new way to detect soft X-ray transients such as XRFs, complemented by rapid onboard multiwavelength follow-up.}
   {We present photometric and spectroscopic observations of XRF~241001A detected by \SVOM, a soft, subluminous, and low-energetic burst located in a poorly populated region of the Amati relation. We investigated the origin of its faint, soft high-energy emission to assess its connection to the long GRB population.}
   {We analyzed the \SVOM/ECLAIRs prompt emission and modeled its afterglow emission from X-ray-to-radio. We present \JWST/NIRSpec and SVOM/VT observations of the associated supernova (SN~2024aiiq), which we modeled with an Arnett radioactive decay component, and we compared its properties with previously detected GRB/SNe.}
   {The event XRF~241001A is located at $z = 0.573$ and has a prompt emission dominated by photons below 20~keV, with a duration of $T_{90} = 3.14$~seconds. Its spectrum is consistent with both thermal and nonthermal models, each implying a low $E_{\rm peak} < 10$~keV and $E_{\rm iso} \sim 8\times10^{49}$~erg. 
   The X-ray-to-radio afterglow modeling favors an origin from a relativistic jet viewed on-axis and supports a nonthermal origin of the prompt emission.
   In the optical, XRF~241001A exhibits an early blue emission, similar to that detected in some extragalactic fast X-ray transients (eFXTs) and inconsistent with the expected synchrotron emission from a relativistic jet.
   The \textit{JWST}/NIRSpec observations firmly established its collapsar origin by revealing an SN Type Ic with broad lines, comparable to SN~1998bw and SN~2025kg-like events near maximum light and with similar properties.}
   {The event XRF~241001A is a soft low-luminosity collapsar event produced by a weak relativistic jet observed on-axis, supporting the view that at least part of the XRF population forms the low-energy soft tail of the long GRB population. Its observation demonstrates the potential of \SVOM/ECLAIRs to probe the soft regime of the high-energy transient population that remains largely unexplored.}
   \titlerunning{SVOM XRF~241001A/SN~2024aiiq}
  \authorrunning{B.~Schneider et al.}
   \maketitle
%

\section{Introduction}
\label{sec:Introduction}
Gamma-ray bursts (GRBs) are among the most energetic transient phenomena in the Universe, releasing isotropic-equivalent energies of $E_{\mathrm{iso}} \sim 10^{48-55}$~erg within seconds \citep{piran2005a,kumar2015a}. 
Historically, GRBs have been divided into two main populations based on their duration (\tgrb) and the spectral hardness of their prompt emission \citep{mazets1981a,kouveliotou1993a}: short bursts ($T_{90} < 2$~s), typically associated with the mergers of compact objects such as neutron stars \citep{eichler1989a,abbott2017a}, and long bursts ($T_{90} > 2$~s), typically linked to the core collapse of massive stars \citep{hjorth2003a,woosley2012a}. 
However, the increasing number of long GRBs attributed to compact object mergers challenges a purely duration-based classification \citep{rastinejad2022a,troja2022a,yang2022a,gompertz2023a}, while events such as GRB~200826A, a short GRB associated with a collapsar, further challenge the traditional GRB classification scheme \citep[e.g.,][]{ahumada2021a,zhang2021a}.
Their brief and intense flashes of high-energy radiation followed by long-lived afterglows across the electromagnetic spectrum, make GRBs unique laboratories for studying relativistic outflows and particle acceleration \citep{kumar2015a}, and they can serve as powerful probes of stellar evolution and the high-redshift Universe \citep[e.g.,][]{savaglio2006a,tanvir2009a,schady2017a,saccardi2023a}.

Long GRBs extend toward softer spectral regimes \citep{amati2002a, ghirlanda2004a}, and depending on their hardness ratio or peak energy ($E_{\rm peak}$), events are classified as X-ray flashes \citep[XRFs;][]{heise2001a,kippen2001a,barraud2003a,pasquale2006a} or X-ray–rich GRBs \citep[XRRs;][]{sakamoto2008a}.
In luminosity space, long GRBs also appear to extend toward lower luminosities, forming a continuum that includes low-luminosity GRBs \citep{liang2007a}. 
First detected by \textit{BeppoSAX} \citep{boella1997a} and \textit{High Energy Transient Explorer 2}  \citep[HETE-2;][]{ricker2003a}, XRFs are characterized by a prompt emission dominated by X-ray photons ($E_{\textrm{peak}}^{\textrm{obs}} \lesssim 30~\mathrm{keV}$) and little or no detectable gamma-ray emission \citep{heise2001a}. 

Several XRFs have been well studied through rapid response and long-term multiwavelength follow-up, including XRF~020903 \citep[e.g.,][]{sakamoto2004a,soderberg2004a}, XRF~030723 \citep[e.g.,][]{fynbo2004a,tominaga2004a,butler2005a}, XRF~050416A \citep[e.g.,][]{holland2006a}, XRF~060218/SN~2006aj \citep[e.g.,][]{campana2006a,ferrero2006a,mazzali2006a,mirabal2006a,modjaz2006a,soderberg2006a,sollerman2006a,pian2006a,toma2007a}, and 
XRF~080109/SN~2008D \citep[e.g.,][]{chevalier2008a,mazzali2008a,soderberg2008a,maund2009a,malesani2009a,modjaz2009a}. 
The current sample indicates that despite their soft spectra, XRFs share many of the observational properties of classical long GRBs, including comparable durations, the presence of X-ray and optical afterglows, associations with supernovae (SNe) Type Ic with broad lines (BLs), and similar host-galaxy environments \citep{sakamoto2008a, bi2018a}.
However, some events such as XRF~080109/SN~2008D were associated with a SN Type~Ib rather than a Ic-BL, suggesting that the origins of XRFs may be diverse \citep[e.g.,][]{chevalier2008a,mazzali2008a}.

The small number of XRFs with extensive early- and late-time multiwavelength observations still limits our understanding of their origin.
Proposed scenarios include GRBs observed off-axis relative to their jet core \citep{yamazaki2002a,zhang2004a,lamb2005a,guidorzi2009a}; intrinsically sub-energetic or soft GRBs, possibly produced by low gamma-ray efficiencies, jets partially or fully choked within the stellar envelope or circumstellar medium, or baryon-loaded GRBs with low Lorentz factors, also referred to as the ``dirty fireball" model \citep[e.g.,][]{dermer1999b,barraud2005a,virgili2009a,nakar2015a,nakar2017a}; supernova shock breakouts \citep{soderberg2008a,chevalier2008a,mazzali2008a,nakar2012a}; or GRBs at very high redshift \citep{heise2001a}.
Each scenario carries different implications. Off-axis events can probe the structure of the jets. \citep{yamazaki2002a,salafia2016a}. Intrinsically soft events provide constraints on the prompt emission mechanisms \citep{dermer1999a,meszaros2002a}. Shock breakout events offer unique insight into the stellar envelope and explosion mechanism \citep{campana2006a,nakar2012a}. Finally, high-redshift GRBs can serve as probes of the early Universe \citep[e.g.,][]{savaglio2006a}.

Since the BeppoSAX and HETE-2 missions, observational progress has been essentially limited by instrumental sensitivity. The \textit{Neil Gehrels Swift Observatory} \citep[\Swift;][]{gehrels2004a}, with its low-energy threshold of 15~keV, has provided partial access to the XRF population, but the lack of coverage below this energy strongly biases detections toward the harder end of the distribution. Missions such as \textit{Fermi} \citep{meegan2009a,atwood2009a} and \textit{Konus-Wind} \citep{aptekar1995a} have extended the coverage to even higher energies but are less sensitive in the soft X-ray domain, where XRFs peak. As a result, the softest GRBs have been underrepresented in recent samples, reflecting the selection effects introduced by instrumental energy thresholds.

The \textit{Space-based multiband astronomical Variable Objects Monitor} \citep[SVOM;][]{wei2016a}, equipped with the wide-field (2~sr) instrument ECLAIRs observing down to 4~keV \citep{godet2014a}, and the \EinsteinProbe mission \citep[\EP;][]{yuan2018a,yuan2022a}, operating in the 0.5--4~keV energy range with a 1.1~sr field of view through its Wide-field X-ray Telescope (WXT), offer a new way to detect and study extragalactic fast X-ray transients (eFXTs), including the XRF population.
The first results from \EP have already provided new and promising insights into the very soft end of the GRB population, paving the way toward a better understanding of the origins of XRFs \citep[e.g.,][]{yin2024a,eyles-ferris2025a,jiang2025a,li2025c,rastinejad2025a,sun2025a,srinivasaragavan2025c}.
In addition, EP260321a/SN~2026gzf highlights the ability of \EP to detect events in which the high-energy signal might arise from a shock breakout with no clear evidence of a successful relativistic jet \citep{chen2026a,martin-carrillo2026a,oconnor2026a,rastinejad2026a,wen2026a,yuan2026a}.

Here we present the detection by SVOM of a soft and faint burst, XRF~241001A \citep{dagoneau2024a,coleiro2024a}, and its multiwavelength analysis, including the detection of the associated SN with the \textit{James Webb Space Telescope} \citep[JWST;][]{gardner2006a}, to investigate its origin and connection to the collapsar GRB population.
The structure of the paper is as follows. In Sect.~\ref{sec:Observation}, we describe the observations and datasets. Section~\ref{sec:Data_analysis} presents the analysis and results. We discuss the implications in Sect.~\ref{sec:Discussion} and summarize our conclusions in Sect.~\ref{sec:Conclusions}.  
Throughout the paper, we adopt a flat $\Lambda$ cold dark matter ($\Lambda$CDM) cosmology from \citet{planckcollaboration2020a}, with $\Omega_{\rm m} = 0.315$, $\Omega_{\Lambda} = 0.685$, and $H_{0} = 67.4$~km~s$^{-1}$~Mpc$^{-1}$. All magnitudes are reported in the AB system, and uncertainties are quoted at the $1\sigma$ level unless otherwise stated. 

\section{Observations and data reduction}
\label{sec:Observation}

\subsection{Gamma-ray detection}
Three months after the launch and during the commissioning phase, ECLAIRs, the gamma-ray coded-mask instrument on board SVOM \citep{godet2014a,godet2026a} triggered on XRF~241001A on 2024 October 1 at 17:08:47.74 UT (hereafter $T_{0}$). The burst was localized at R.A. = $20.42^\circ$, Dec = $-43.53^\circ$ (J2000), with a 90\% confidence radius of 11$^\prime$ including a 2$^\prime$ systematic added in quadrature \citep{dagoneau2024a,coleiro2024a}. The platform did not slew because automated slewing was not yet activated.

\begin{figure}[t]
    \centering
    \includegraphics[width=\hsize]{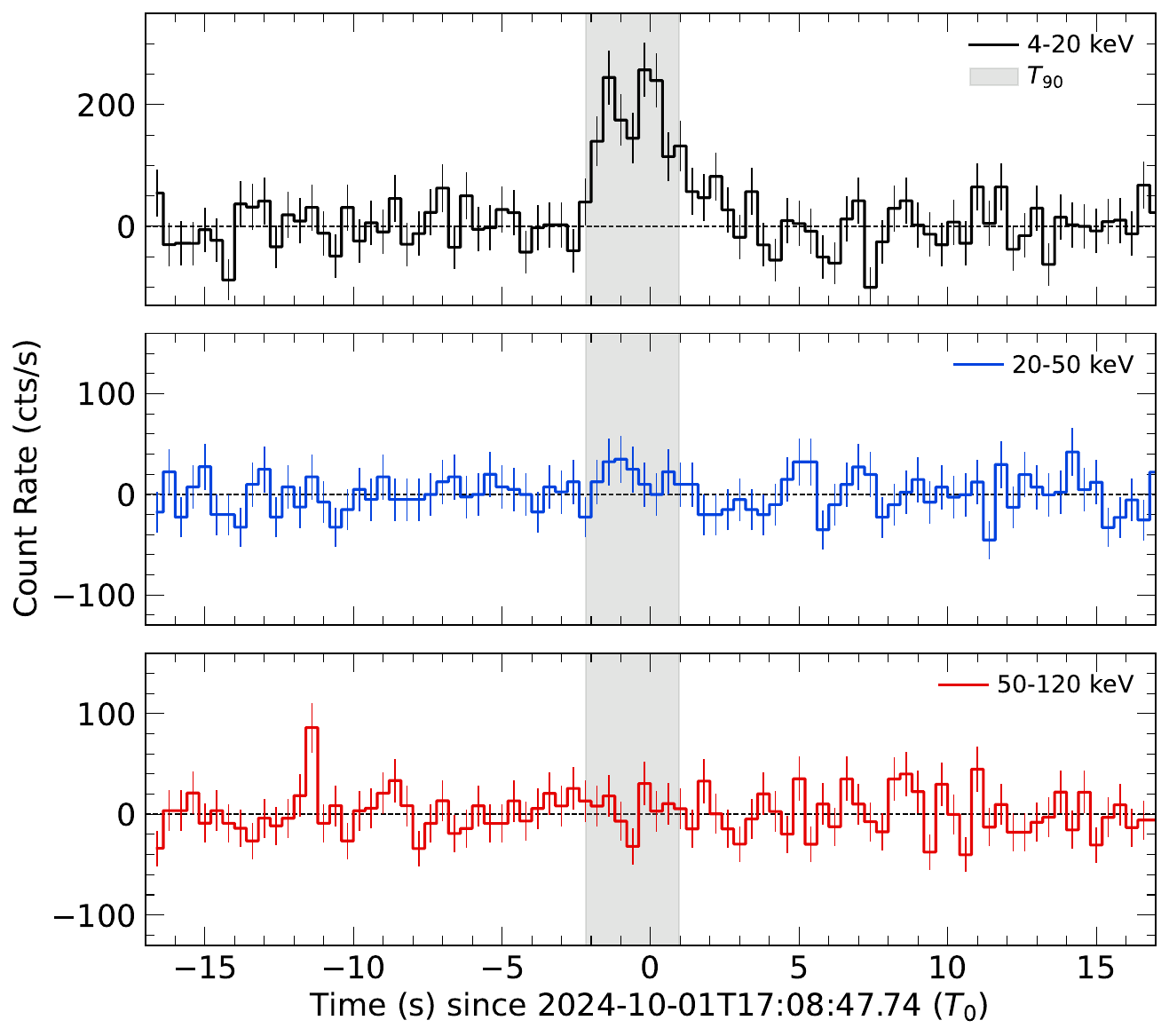}
    \caption{Light curve of the prompt emission of XRF~241001A detected by SVOM/ECLAIRs. The background-subtracted signal is centered on the trigger time $T_{0} = \mathrm{2024\text{-}10\text{-}01T17{:}08{:}57.74}$ UT and shown in three energy bands: 4--20~keV (top, black), 20--50~keV (middle, blue), and 50--120~keV (bottom, red). The gray shaded region indicates the $T_{90}$ interval of the burst.}
    \label{fig:eclairs_lc}
\end{figure}

\begin{figure}[t]
    \centering
    \includegraphics[width=\hsize]{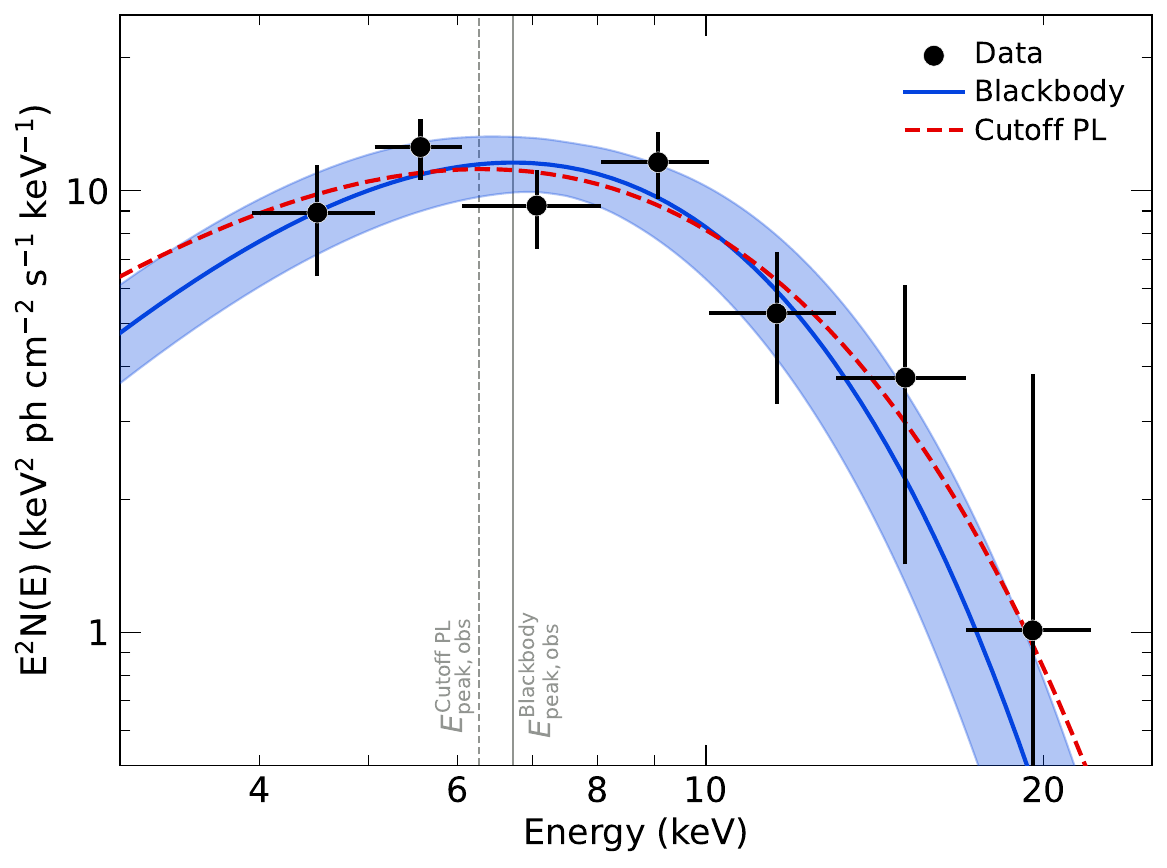}
    \caption{Prompt emission spectrum of XRF~241001A observed with SVOM/ECLAIRs. The data are shown in black. The best-fit blackbody model is shown as a blue solid line, and the cutoff power-law model is shown as a red dashed line. The observed peak energies of the cutoff power-law and blackbody models are marked by gray dashed and solid lines, respectively. For clarity, only the 1$\sigma$ uncertainty of the blackbody model is shown.}
    \label{fig:eclairs_spec}
\end{figure}

Using X band event by event data, the light curve was produced using only the pixels of the detector plane that were illuminated by the source, which for this burst corresponds to 29.6\% of the pixels (Fig.~\ref{fig:eclairs_lc}). This method minimizes the background noise and therefore maximizes the signal-to-noise ratio (S/N). 
The spectrum was extracted with the ECLAIRs pipeline (ECPI, version 1.18.3), which executes a sequence of processing stages to generate the files containing the spectral data (Fig.~\ref{fig:eclairs_spec}).

First, the pipeline prepared the raw data to produce higher-level scientific products, such as spectra. Orbital and altitude parameters, along with housekeeping data, were used to determine the good time intervals (GTIs). 
The energy of the events was subsequently derived through a pixel-dependent linear calibration function that incorporates individual gain and offset parameters. The calibrated event was then associated with a unique Pulse Invariant (PI) channel by identifying the energy interval that corresponds to its value in the reference table.
Based on the GTIs, the pipeline selected the calibrated events within these intervals to construct raw detector images. Uniformity and background maps were generated, while efficiency and nonuniformity correction maps were applied to account for instrumental variations. The resulting detector images, produced over 8 user-defined energy bands (see Appendix~\ref{app:additional_materials}), were corrected for background and Earth occultation.
Then, the reconstruction of sky images was performed through a deconvolution procedure applied to the corrected detector images in each energy band. 

No additional sources were detected within the sky images using a S/N threshold of 10. 
Finally, the pipeline constructs a shadowgram model of the burst using the known mask pattern and the source position derived from the sky images. This model, including both the burst and background contributions, is then fitted to the detector map, allowing the source flux to be extracted in each energy band.\footnote{The response matrix file ECL-RSP-RMF\_20220515T103600 and the ancillary response file ECL-RSP-ARF\_20220515T104100 were used in the analysis.} The background is modeled with a second-degree polynomial.
More details on the ECLAIRs data reduction can be found in \cite{cordier2025a} and \cite{goldwurm2026a}.

\subsection{X-ray follow-up}
We performed two follow-up observations with the \Swift/XRT instrument \citep{burrows2005a}, starting at $T-T_{0} = 1.81$~hours and 4.3 days for 3.2~ks and 4.2~ks of exposures, respectively \citep{osborne2024a}. 
We also obtained one target-of-opportunity (ToO) observation with the \EP Follow-up X-ray Telescope \citep[\EP/FXT;][]{chen2025b} starting at $T-T_{0} = 12.4$~hours for a total exposure of $\sim$6~ks \citep{turpin2024a}.
An uncatalogued X-ray source was detected by both \Swift/XRT (first epoch) and \EP/FXT observations inside the ECLAIRs error radius at a distance of 6.65$^\prime$ from the ECLAIRs derived position. The source was no longer detected in the late time \Swift/XRT observation and showed a clear sign of fading along the different epochs. This allowed us to secure the identification of this source as the X-ray afterglow of XRF~241001A. The \Swift/XRT enhanced position is R.A. = 20.5529$^\circ$ and Dec = $-43.4752^\circ$ (J2000) with a 90$\%$ confidence level error radius of $\sigma = 3.6^{\prime\prime}$ and is consistent with the \EP/FXT position.

The \textit{Swift}/XRT spectrum\footnote{Associated with the RMF file swxpc0to12s6$\_$20210101v016.rmf and the ARF created from the xrtmkarf routine v0.6.4} of the first epoch was retrieved from the UK \Swift\ Science Data Centre \citep{evans2007a,evans2009a} using the \Swift/XRT Data Analysis Software v3.7.0 (Heasoft v6.32).
The \EP/FXT (A and B) spectra were processed following the standard data reduction procedure described in the \EP/FXT Data Analysis Software Package (FXTDAS v1.20). 
Due to the low S/N of the spectra, we grouped the energy bins to ensure at least one count per bin while preserving sufficient spectral resolution for fitting (see Sect.~\ref{sec:xrayanalysis}). The grouping was performed using the \textit{grppha} ftools routine and the data were fitted using the {\sc xspec} package v12.14.1 \citep{arnaud1999a}. 

\subsection{Optical follow-up}
The Ultra-Violet/Optical Telescope (UVOT; \citealt{roming2005a}) on board \Swift started observing the field of XRF~241001A at 1.81 hours after the trigger. A fading optical source was detected in the $u$ band, spatially coincident with both the \Swift/XRT afterglow position and the optical counterpart identified by LCOGT \citep{izzo2024a} at R.A. = 20.553129$^\circ$ and Dec = $-43.475542^\circ$ (J2000), with an uncertainty of 0.5$^{\prime\prime}$. We retrieved the magnitudes reported by \cite{breeveld2024a} and converted them to the AB magnitude system using \cite{breeveld2011a}. At the second epoch, 4.3~days after the trigger, the source had faded below the sensitivity of \Swift/UVOT and was no longer detected. We derive a 3$\sigma$ upper limit using the \Swift\ software (v5.9) provided as part of the HEASOFT package (v6.36). 

The Visible Telescope (VT) is the optical telescope on board SVOM \citep{fan2020a,qiu2026a}. It has an effective aperture of 44~cm and offers a $26^\prime \times 26^\prime$ field of view with a pixel scale of $0.76^{\prime\prime}$. The telescope operates simultaneously in two channels, VT\_B and VT\_R, covering wavelength ranges of 400–650~nm and 650–1000~nm, respectively.
XRF~241001A was observed with VT during seven epochs via Target Of Opportunity (ToO) observations between October 2, 2024, and October 14, 2024 \citep{qiu2024a}. The exposure time for each individual frame was 20~seconds and the total exposure time was different for each epoch (Table~\ref{table:photometry}). 
The data were processed with a custom-developed code based on the IRAF package using standard procedures, including bias subtraction, dark correction, and flat-field correction. For each epoch and filter, the individual frames were stacked to improve the S/N. Photometry was performed in the AB magnitude system with an aperture radius of 1.5~pixels (i.e., $1.14^{\prime\prime}$). A detailed description of the reduction procedure is presented in \citet{li2026a}.  

We observed the field of the XRF~241001A with the Sinistro imager mounted on the 1-meter telescope of the Las Cumbres Observatory Global Telescope (LCOGT) network \citep{brown2013a} located at the South African astronomical observatory in Sutherland. Observations started on October 1, 2024, at 18:40:35 UT, approximately 1.53~hours after the GRB trigger \citep{izzo2024a}. We captured a sequence of 3$\times$120 s exposures in the $r^{\prime}$ SDSS filter and 5$\times$120 s exposures in the $z$ PS1 filter. 
We obtained reduced frames from the LCOGT web archive and then stacked them using SWarp \citep{bertin2010a}. The final resulting images show poor tracking of the telescope, resulting in a small drift for all stars in the stacked images. 
Similarly to the X-shooter $z$ band images, the LCOGT $z$ band images were affected by ``fringes'' and were corrected using the method described in Sect.~\ref{ssect:x-shooter}.
In the final stacked images, we identified an uncatalogued source within the Swift/XRT error circle position. Aperture photometry was performed using the SkyMapper DR4 catalog \citep{onken2024a} as a photometric reference, and the resulting magnitudes are reported in Table~\ref{table:photometry}.

The Global Rapid Advanced Network Devoted to the Multi-messenger Addict \citep[GRANDMA;][]{antier2020b} started observations with the Télescope à action rapide pour les objets transitoires Réunion \citep[TAROT/TRE;][]{klotz2019a} at Les Makes observatory 43~min post $T_{0}$ without filter during 19$\times$120~s exposures, and finally by the iTelescope T72 from the Kilonova-catcher (KNC) program 11.43~hours post $T_{0}$ in Johnson $R$ with exposures of 18$\times$180 s \citep{turpin2024b}. We used the STDPipe pipeline and its web interface, to perform forced photometry \citep{karpov2025a}. The photometry of the images in Bessel $R$ and clear was calibrated using the SkyMapper DR4 catalog \citep{onken2024a} and converted in the $r$ band to be consistent with the rest of the measurements, using the color term approach \citep{karpov2025a}. In our final stacked images, the source is not detected in any of the epochs, and we report a 3$\sigma$ upper limit in the photometric Table~\ref{table:photometry}. 

We obtained $g^{\prime}$ and $r^{\prime}$ SDSS band observations of the source using the Low Dispersion Survey Spectrograph 3C (LDSS-3C; \citealt{stevenson2016a}), an optical imaging spectrograph mounted on the Magellan Clay telescope \citep{shectman2003a} about 69~days after the GRB. 
Observations began under good conditions, with 0.6\arcsec{} seeing, at 05:00:26 UT on 2024-12-09 and consisted of 4$\times$300~s exposures in g$^{\prime}$ and 5$\times$300~s exposures in r$^{\prime}$. 
Data reduction was performed using custom Python scripts, including bias subtraction, flat-fielding, and stacking. Astrometric calibration was carried out using \textit{Astrometry.net}\footnote{\url{https://astrometry.net}} software, and the photometry was calibrated with reference to the SkyMapper DR4 catalog \citep{onken2024a}.
In our stacked image, no source is detected at the afterglow position and we compute a 3$\sigma$ upper limit for each band (Table~\ref{table:photometry}).

\subsection{Optical/NIR spectroscopy}
\label{ssect:x-shooter}
\begin{figure*}[t]
\centering
    \includegraphics[width=17cm]{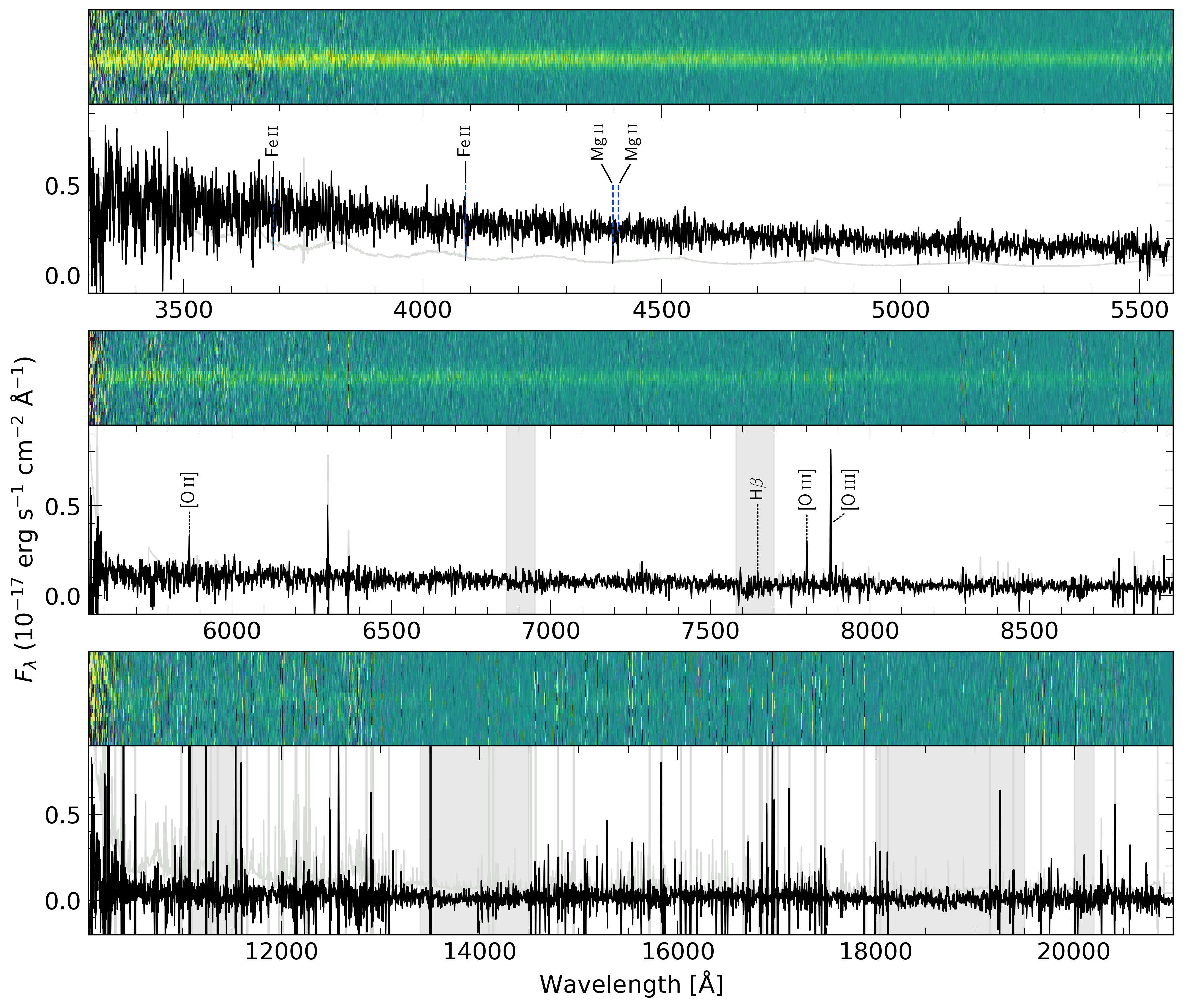}
    \caption{Spectra from VLT/X-shooter of XRF~241001A $(z = 0.5728 \pm 0.0001)$ obtained at 0.58~days after the trigger. The UVB, VIS, and NIR arms are displayed (from top to bottom). In each panel, the upper subpanel shows the 2D spectrum, while the lower subpanel presents the 1D extracted spectrum.
    Absorption and emission lines are highlighted and labeled as blue dashed and black dotted vertical lines, respectively.
    The 1D spectra have been smoothed using a Savitzky–Golay filter to enhance the visibility of the lines, and the associated uncertainties are shown as a gray line. Gray shaded regions mark wavelength intervals strongly affected by telluric absorption.}
    \label{fig:x-shooter}
\end{figure*}

Following the detection of the optical afterglow reported by \cite{izzo2024a}, we carried out spectroscopic observations under the ESO program ID 110.24CF.015 (PIs: Tanvir, Malesani, Vergani) using the X-shooter spectrograph \citep{vernet2011a} mounted on the ESO Very Large Telescope (VLT) at Cerro Paranal, Chile. X-shooter provides simultaneous spectral coverage in the ultraviolet-blue (UVB; 300--560~nm), visible (VIS; 550--1020~nm), and near-infrared (NIR; 1020--2100~nm) arms, with resolving powers of R $\sim$ 5400, 8900, and 5600, respectively. The 0.9\arcsec{}JH slit with the $K$ band blocking filter was used to minimize the amount of light scattered from the $K$ band and improve the sky background in the NIR spectrum. 
The observations consisted of 4$\times$1200 s and started at 06:31:53 UT on 2024-10-02 (13.39~hours after the SVOM trigger). The spectra were obtained under good conditions with a seeing of 0.7\arcsec{} and an airmass of 1.3 (Fig.~\ref{fig:x-shooter}). The instrument was operated in nodding mode using an ABBA sequence to improve the sky subtraction, particularly in the NIR. 

Data reduction was performed in STARE mode using the standard ESO X-shooter pipeline \citep{goldoni2006a, modigliani2010a}, which includes bias and dark subtraction, flat-field correction, wavelength calibration using arc lamps, and flux calibration based on observations of standard stars taken on the same night. The individual reduced exposures of each arm were then combined, and the residual sky background was subtracted \citep{selsing2019a}. 
Prior to the X-shooter spectroscopic observations, the source was imaged using the acquisition and guiding camera in the $g^{\prime}$, $r^{\prime}$, and $z^{\prime}$ SDSS filters with exposures of 3$\times$60 s each (Table~\ref{table:photometry}). 
Data reduction was conducted using custom Python scripts and included bias subtraction, flat correction, and stacking. The $z$ band images were affected by ``fringes,'' which were corrected by creating a master fringe frame computed with the mean of the dithered images. The resulting master image is scaled and subtracted from science images after bias and flat-field correction. Photometric calibration was performed using nearby stars from the SkyMapper DR4 catalog \citep{onken2024a} and the magnitudes were computed using the aperture photometry method (Table~\ref{table:photometry}).

The calibrated spectra may be subject to various effects that result in systematic errors in the flux determination, such as differences between the observing conditions of the target and those of the spectrophotometric standard stars, or slit losses due to seeing.
The magnitudes obtained from the acquisition and guiding camera were used to re-normalize the flux. The best scaling factor was determined by minimizing the difference between the observed $g^{\prime}$, $r^{\prime}$, and $z^{\prime}$ magnitudes, and synthetic magnitudes were computed from the scaled spectrum. We found a scaling factor of 2.1 between the observed photometry and spectrum.

The James Webb Space Telescope \citep[JWST;][]{gardner2006a} observed the field with the Near Infrared Spectrograph \citep[NIRSpec;][]{jakobsen2022a} under program 6133 (PI: Gompertz). The PRISM/CLEAR configuration was used and provided low-resolution spectroscopy (${\rm R} \sim 100$) in the 0.6 to 5.3~$\mu$m wavelength range. The Fixed Slit Spectroscopy mode with the S400A1 slit ($0.4\arcsec \times 3.65\arcsec$) was set at the best known afterglow position. The observations started on 2024-10-23 at 02:51:23.72 UT, approximately 21.4~days after the trigger (13.6~days in rest-frame), close to the expected SN peak \citep{cano2017a} for a total exposure time of 1.75 hours \citep{gompertz2024a}. The level 3 science products \citep{bushouse2023a}, including the 1D and 2D spectra, were retrieved from the Mikulski Archive for Space Telescopes (MAST).

\begin{figure*}[t]
    \begin{subfigure}[b]{0.49\hsize}
        \centering
        \includegraphics[width=\hsize]{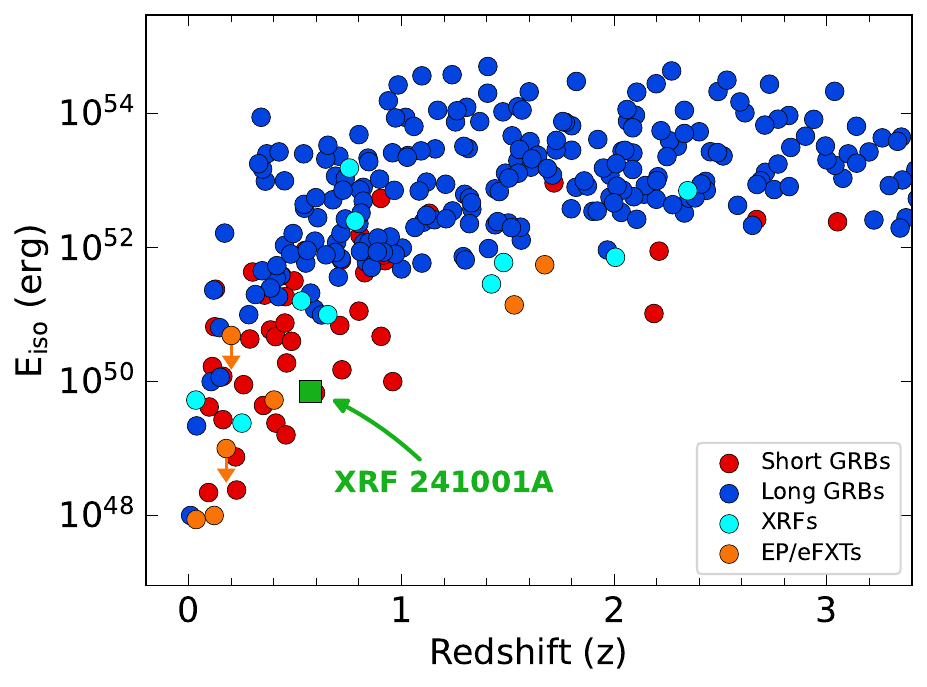}
    \end{subfigure}
    \hfill
    \begin{subfigure}[b]{0.49\hsize}
        \centering
        \includegraphics[width=\hsize]{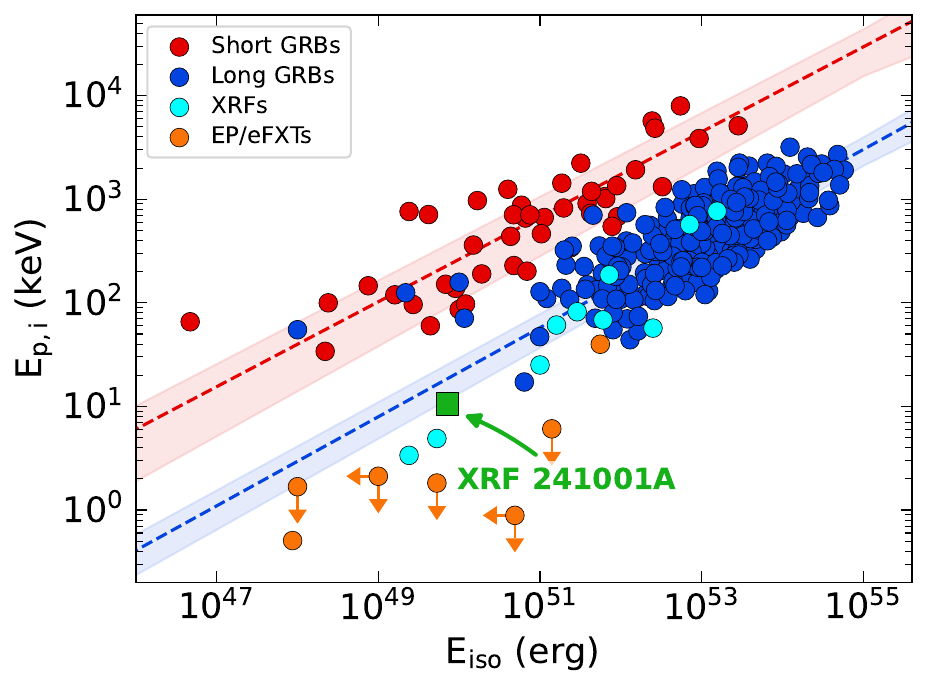}
    \end{subfigure}
    \caption{Prompt emission properties of XRF~241001A compared to the GRB sample in \cite{minaev2020a} and eFXTs detected by \EP from \cite{dai2026a,jiang2025a,li2025c,sun2025a,srinivasaragavan2025c,cotter2026a,yuan2026a}. The red circles represent short GRBs, the blue circles are for long GRBs, the cyan circles are for XRFs, and the orange circles are for \EP/eFXTs. XRF~241001A is shown as a green square. \textit{Left panel:} Isotropic energy ($\mathrm{E}_{\mathrm{iso}}$) from 1 to 10 000~keV as a function of the redshift. \textit{Right panel:} Amati plot, $\mathrm{E}_{\mathrm{p,i}}$ against $\mathrm{E}_{\mathrm{iso}}$. The best-fit model of the Amati relation is visible in red for short GRBs and in blue for long GRBs. The 1$\sigma$ error is indicated as a shaded area.}
    \label{fig:eclairs_amati}
\end{figure*}

\subsection{Radio observations}
The field was observed with the Australia Telescope Compact Array (ATCA; \citealt{frater1992a,wilson2011a}) under project CX583 (PI: Leung). We carried out observations on 6 and 19 October 2024 (5 and 18~days post-burst) with the 5.5/9~GHz receiver and the array in 6A configuration\footnote{\url{https://www.narrabri.atnf.csiro.au/operations/array_configurations/configurations.html}} \citep{leung2024a}.
We reduced the data following standard {\sc Miriad} procedures \citep{sault1995a}. 
To calibrate the flux-density scale and the bandpass, we used B1934$-$638, and to track how the complex gains change over time, we periodically observed the compact and bright gain calibrator B0104$-$408. 
We de-convolved and imaged the data using the multifrequency synthesis CLEAN algorithm \citep{hogbom1974a,clark1980a,sault1994a}. 
For detections, we fit a point-source model to obtain a flux-density measurement, while for a non-detection we take the $3\sigma$ upper limit, where $\sigma$ is the rms sensitivity of the residual image in the region where the target is located (Table~\ref{table:radio}). 

\section{Data analysis and results}
\label{sec:Data_analysis}

\subsection{Redshift measurement}
\label{ssect:redshift}
Our X-shooter spectrum obtained at 0.58~days after the ECLAIRs trigger reveals a continuum over the whole range of the spectrum, albeit only faintly in the NIR arm. In the blue part of the UVB spectrum, multiple absorption lines were detected and identified as low-ionization transitions produced by Fe~\textsc{ii} $\lambda$2344, 2600, and Mg~\textsc{ii} $\lambda$2796, 2804. The Fe~\textsc{ii} $\lambda$2344 feature is detected with a relatively low S/N.
We fitted these lines with a Voigt profile using the {\sc VoigtFit} Python package \citep{krogager2018a} and measured a redshift of $z_{abs} = 0.5728 \pm 0.0001$, consistent with the prompt value reported by \cite{palmerio2024a}. 

In the spectrum of the VIS arm, multiple emission lines were also identified as being due to [O \textsc{ii}] $\lambda$$\lambda$3726, 3729, H$\beta$, [O \textsc{iii}] $\lambda$4959, and [O \textsc{iii}] $\lambda$5007 of the underlying host galaxy. A hint of a faint H$\alpha$ line is also marginally visible in the 2D spectrum at low significance level (<3$\sigma$). From a Gaussian fit of these lines using the {\sc lmfit} Python package \citep{newville2025a}, we inferred a redshift of $z_{em} = 0.5729 \pm 0.0001$. The 1D and 2D X-shooter spectra of the three arms are shown in Fig.~\ref{fig:x-shooter}, with labeled absorption and emission lines.

\subsection{Prompt emission analysis} 
Figure~\ref{fig:eclairs_lc} shows the prompt emission of XRF~241001A detected by \SVOM/ECLAIRs in three different energy bands (4--20, 20--50 and 50--120~keV). It indicates that the emission is predominantly observed below 20~keV, with no significant emission above this energy, placing the event in the XRF regime. We also examine the data obtained with the Gamma Ray Monitor (GRM; \citealt{dong2010a}) on board SVOM, but no significant signal is detected during the ECLAIRs trigger.

In the 4--20~keV band, we measured a duration of $T_{90} = 3.14\pm 0.18$~s using the \textsc{battblock} function (v1.18). The time-averaged spectrum over the $T_{90}$ interval was extracted in the 4--20~keV energy range.
The prompt emission spectrum can be fitted by both nonthermal and thermal models (Fig.~\ref{fig:eclairs_spec}). Nonthermal models, such as the cutoff power-law and broken power-law models, provide a good fit of the spectrum, but their parameters are not constrained due to the faintness of the signal. The cutoff power-law model ($\chi^2/\mathrm{d.o.f.} = 3.43/4$) returns a $E_{\rm peak} = 6.27^{+0.87}_{-1.07}$~keV and a flux $1.96^{+0.07}_{-0.08} \times 10^{-8} \ \mathrm{erg} ~\mathrm{cm}^{-2}~ \mathrm{s}^{-1}$ in 4--20~keV. A power-law model gives a poor fit to the data ($\chi^2/\mathrm{d.o.f.} = 8.45/5$). Alternatively, a thermal blackbody model can fit the data well ($\chi^2/\mathrm{d.o.f.} = 3.72/5)$ with a temperature of $kT = 1.72^{+0.16}_{-0.15}$~keV, corresponding to $E_{\rm peak} = 6.73^{+0.61}_{-0.57}$~keV (using $E_{\rm peak} = 3.92\,kT$) and a flux of $1.89^{+0.11}_{-0.40} \times 10^{-8} \ \mathrm{erg} ~\mathrm{cm}^{-2}~ \mathrm{s}^{-1}$ in 4--20~keV. 
The time-integrated spectral results for the tested models are reported in Table~\ref{tab:xrf241001a_prompt} and the 1~s peak analysis is summarized in Table~\ref{tab:xrf241001a_1speak}.

All models (nonthermal or thermal) provide $E_{\rm peak}<10$~keV, well below the $\sim$30~keV threshold typically used to classify XRF and support its identification as an XRF. 
The classification method, such as the one proposed by \citet{sakamoto2008a} for the \Swift/BAT energy range based on the fluence ratio $S(25\text{–}50~\mathrm{keV})/S(50\text{–}100~\mathrm{keV})$, cannot be reliably applied to XRF~241001A due to the absence of significant emissions detected above 20~keV. A classification method adapted to the \SVOM/ECLAIRs energy bands will be investigated in a future work with a larger sample of events.

\begin{figure*}[t]
    \begin{subfigure}[b]{0.49\hsize}
        \centering
        \includegraphics[width=0.94\hsize]{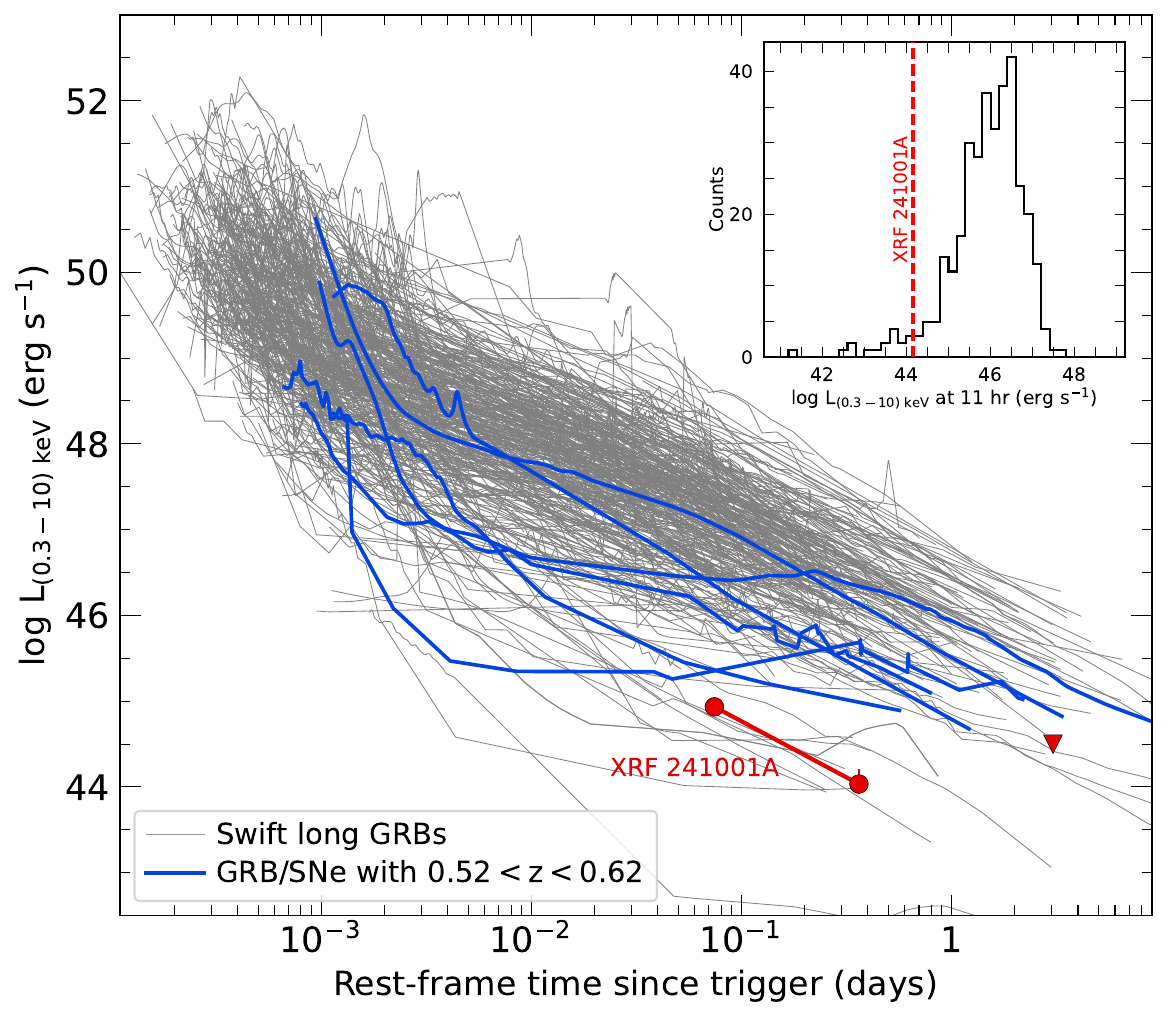}
    \end{subfigure}
    \hfill
    \begin{subfigure}[b]{0.49\hsize}
        \centering
        \includegraphics[width=\hsize]{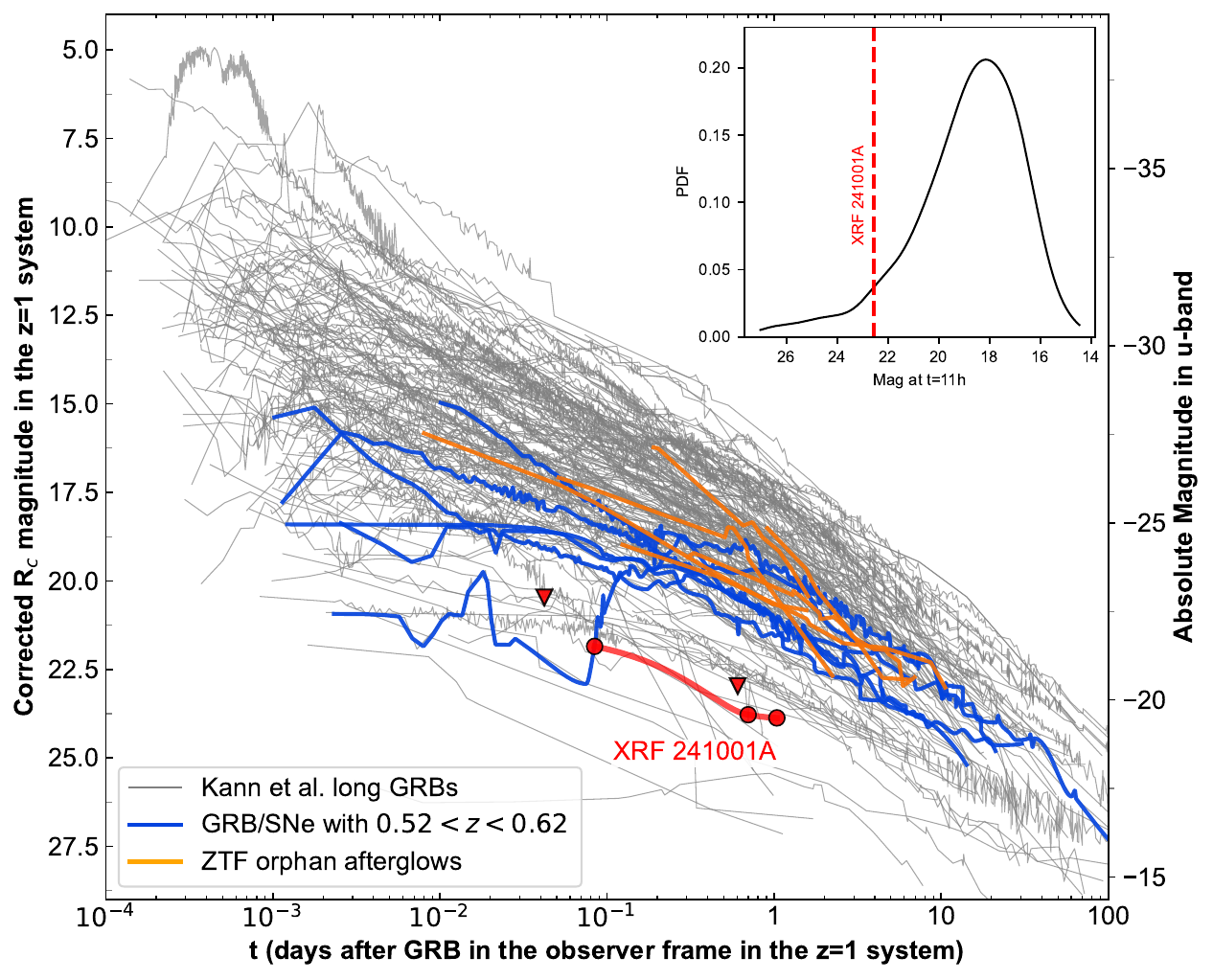}
    \end{subfigure}
    \caption{Comparison of the XRF~241001A afterglow light curve with those of previously detected long GRBs.
    \textit{Left panel:} X-ray luminosity in the 0.3–10~keV band as a function of rest-frame time for XRF~241001A (red) and long GRBs observed by \Swift/XRT (gray). The GRBs with an associated SN at $0.52 < z < 0.62$ are highlighted in blue. The inset plot shows the distribution of the X-ray luminosity at 11~hr in rest-frame after the trigger for the \Swift/XRT GRB sample and for XRF~241001A (red dashed line).
    \textit{Right panel:} $R_\mathrm{c}$ band AB magnitude as a function of time (Kann plot) after shifting all light curves to a common redshift of $z = 1$. Long GRBs from \cite{kann2010a,kann2011a} are shown in gray, GRB/SNe at $0.52 < z < 0.62$ in blue, XRF~241001A in red, and ZTF orphan afterglows in orange \citep{ho2020a, ho2022a, li2025a, perley2025a, srinivasaragavan2025a}. The inset panel displays the probability density function (PDF) of long GRB magnitudes at 11~hr post-trigger, with XRF~241001A marked by a dashed red line. Upper limits are shown as red triangles in both panels.}
    \label{fig:AG_comparison}
\end{figure*}

Using spectroscopic observations of the afterglow emission, the redshift of the burst was determined to be $z = 0.5728$ (Sect.~\ref{ssect:redshift}). Using this value and the best-fit models, we derive an isotropic peak luminosity of $L_{\rm iso} = 4.48^{+1.05}_{-0.88} \times 10^{49}$erg\,s$^{-1}$ and an isotropic energy of $E_{\rm iso} = 8.95^{+2.10}_{-1.76} \times 10^{49}$~erg for the cutoff power-law model and we find consistent values for the blackbody model (see Table~\ref{tab:xrf241001a_prompt}).
Both models establish it as a subluminous burst given its redshift and place it among the least energetic GRBs detected at comparable distances (left panel of Fig.~\ref{fig:eclairs_amati}). 
In the Amati diagram, XRF~241001A is located in a poorly sampled region of low $E_{\rm iso}$ and low $E_{\rm peak}$, extending the correlation of the long GRB population toward fainter and softer events, similar to other XRFs such as XRFs~060218 and 020903 (right panel of Fig.~\ref{fig:eclairs_amati}).

\subsection{X-ray afterglow analysis}
\label{sec:xrayanalysis}
Given the low number of counts and the binning of one count per energy bin in the \Swift/XRT and \EP/FXT spectra, the data are treated as Poisson-distributed rather than Gaussian, and we use the \textit{C-statistic} \citep{cash1979a} to perform the spectral fitting. We first check that there is no significant spectral evolution between the \Swift and \EP epochs by fitting the 0.3--10 keV spectra individually with a simple absorbed power law model \textsf{$Tbabs \times zTbabs \times pow$} using the abundance table for the X-ray interstellar absorption described in \cite{wilms2000a}. $Tbabs$ is fixed at the Galactic value $N^{\rm Gal}_{H,X} = 1.28\times 10^{20}\,\mathrm{cm}^{-2}$ \citep{bekhti2016a}. 
For the \Swift/XRT spectrum, we measured $N^{\rm host}_{H,X}=0.57^{+0.54}_{-0.36} \times 10^{22}\,\mathrm{cm}^{-2}$ and $\Gamma = 1.70^{+0.45}_{-0.37}$. On the other hand, for the \EP/FXT spectrum, which has a low S/N, we only let $N^{\rm host}_{H,X}$ vary within the 1$\sigma$ error bar of the $N^{\rm host}_{H,X}$ value found from the \textit{Swift}/XRT spectral analysis. We found $N^{\rm host}_{H,X}=~0.61^{+0.32}_{-0.28} \times 10^{22}\,\mathrm{cm}^{-2}$ and $\Gamma = 2.30^{+0.65}_{-0.60}$. Compared to the \Swift/XRT spectrum, the \EP/FXT spectrum does not show a significant spectral evolution, with photon indices compatible within their 1$\sigma$ errors.
Assuming the absence of spectral evolution, we also performed a joint spectral fit to better constrain the parameters. From the joint fit, we measure $N^{\rm host}_{H,X} = 0.52^{+0.41}_{-0.30} \times 10^{22}\,\mathrm{cm}^{-2}$ and $\Gamma = 1.80^{+0.37}_{-0.30}$. The results of our spectral analysis and the estimated 0.3--10 keV fluxes are summarized in Table~\ref{tab:x-ray_spec}.

We retrieved the \Swift/XRT light curves \citep{evans2007a, evans2009a}, redshifts, and the photon indices of the time-averaged spectrum provided by the \Swift\ Burst Analyzer \citep{evans2010a} using the {\sc swifttools.ukssdc} package\footnote{\url{https://www.swift.ac.uk/API}} for GRBs detected between December 2004 and November 2025. 
The unabsorbed X-ray light curves are converted into X-ray luminosities in the 0.3–10~keV band following \cite{schulze2014a} 
Each X-ray light curve was smoothed using a LOWESS \citep[Locally Weighted Scatterplot Smoothing,][]{cleveland1979a} algorithm to improve visual clarity while preserving the global temporal decay behavior of the afterglow.

In the left panel of Fig.~\ref{fig:AG_comparison}, the X-ray light curve of long \Swift/XRT GRBs ($T_{90} > 2$~s) is shown in comparison with that of XRF~241001A. GRBs with an associated SN in the redshift range $0.52 < z < 0.62$ \citep[GRBs~050525A, 060729, 081007, 090618, 101219B,][]{finneran2025a} are highlighted in blue.
XRF~241001A is located at the faint end of the luminosity distribution of GRBs previously detected by \Swift/XRT. 
We estimate the rest-frame 11-hour X-ray luminosity by locally fitting a single power-law decay to each light curve and evaluating the fit at the corresponding epoch. We note that, in a limited number of cases, this simple model might not fully capture the true temporal decay behavior, which can include rebrightenings or breaks.
We find that XRF~241001A lies among the faintest $\sim$5\% of the sample (see the inset plot in the left panel of Fig.~\ref{fig:AG_comparison}), corresponding to an exceptionally low X-ray afterglow luminosity.

\subsection{Optical afterglow analysis}
\label{ssection:optical_afterglow_analysis}
We used the spectroscopic data obtained from the VLT/X-shooter normalized to the observed photometry (Sect.~\ref{ssect:x-shooter}) and JWST/NIRSpec instruments to supplement the optical light curves derived from imaging.
From the VLT/X-shooter spectrum, we computed synthetic magnitudes in the \Swift/UVOT $U$, SDSS $i^{\prime}$, VT\_B and VT\_R filters using {\sc pyphot} \citep{fouesneau2025a}. The uncertainties were estimated via the Monte Carlo method by perturbing the spectrum within its flux errors, and a 5\% systematic term was added to account for various sources of uncertainty in the reduction and flux-calibration processes.
We note that the X-shooter UVB arm covers only about 80\% of the \Swift/UVOT $U$ filter, which might lead to an under- or over-estimation of the corresponding synthetic magnitude depending on the intrinsic spectral energy distribution (SED) of the source. To mitigate this effect, we fit the combined spectrum from UVB to NIR with a power-law model and assume negligible host-galaxy dust extinction, as suggested by the afterglow modeling (see Sect.~\ref{ssection:agmodeling}). The best-fit model provides a spectral index of $\beta = -0.4 \pm 0.1$ (see also Sect.~\ref{ssection:blue_emission}) that we use to extrapolate the emission at the blue low-end and derive the synthetic magnitude for the \Swift/UVOT $U$ band (Table~\ref{table:photometry}).

Similarly, we extracted synthetic magnitudes from the JWST/NIRSpec spectrum in the SDSS $i^{\prime}$, SDSS $z^{\prime}$, and VT\_R filters. 
The limited coverage toward the blue end of the JWST/NIRSpec spectrum requires extending the signal below 0.6~µm to compute synthetic magnitudes in the SDSS $g^{\prime}$, SDSS $r^{\prime}$, and VT\_B bands. However, the spectrum shows an emission consistent with the presence of an emerging SN component (Sect.~\ref{ssection:SupernovaAnalysis}). In contrast to an afterglow-dominated spectrum, this thermally dominated emission requires particular caution when extrapolating. For this reason, only the bands fully covered by our spectrum ($i^{\prime}$, $z^{\prime}$, and VT\_R) are included in our modeling to be conservative.
The optical magnitudes used in the analysis are listed in Table~\ref{table:photometry} and are corrected for Galactic dust extinction, assuming $A_V^{\rm Gal} = 0.03$~mag \citep{schlafly2011a}.

In the right panel of Fig.~\ref{fig:AG_comparison}, we compare the $r$ band light curve of XRF~241001A with the sample of GRB afterglows compiled by \cite{kann2010a,kann2011a}. All light curves were shifted to a common redshift of $ z = 1$ following the prescription of \cite{kann2006a}. The GRB/SNe in the redshift range $0.52 < z < 0.62$ are highlighted in blue. Similarly to the X-ray light curve, XRF~241001A is found at the faint end of the distribution and is consistent with being among the faintest $\sim$6\% of events in the sample.

The faintness of the prompt emission detected by ECLAIRs motivates a comparison with ZTF orphan afterglows, which are optical transients exhibiting a typical afterglow-like decay but with no associated GRB detection. Specifically, we retrieve the $r$ band light curves of AT2019pim, AT2020blt, AT2021any, AT2021lfa, AT2023lcr, and AT2023sva from \cite{ho2020a, ho2022a, li2025a, perley2025a, srinivasaragavan2025a}. The data are shifted to a common redshift of $z = 1$, assuming a typical optical spectral slope of $\beta_{\rm O} = 0.6$ \citep{kann2006a, kann2010a}. Owing to the lack of reliable estimates of the host galaxy extinction, we correct only for Galactic extinction, noting that any additional host extinction would increase the intrinsic optical luminosities.
We find that the optical afterglow of XRF~241001A is significantly fainter than the ZTF orphan afterglows, differing by at least $\sim$2 magnitudes at 11~hr after the trigger. 
This offset might indicate that the orphan afterglows belong to a different population than XRF~241001A, or reflect selection effects in the ZTF survey, which preferentially detects intrinsically bright afterglows \citep{ho2022a,sarin2022a}.

\subsection{Early blue optical emission}
\label{ssection:blue_emission}
\begin{figure}[t]
    \centering
    \includegraphics[width=\hsize]{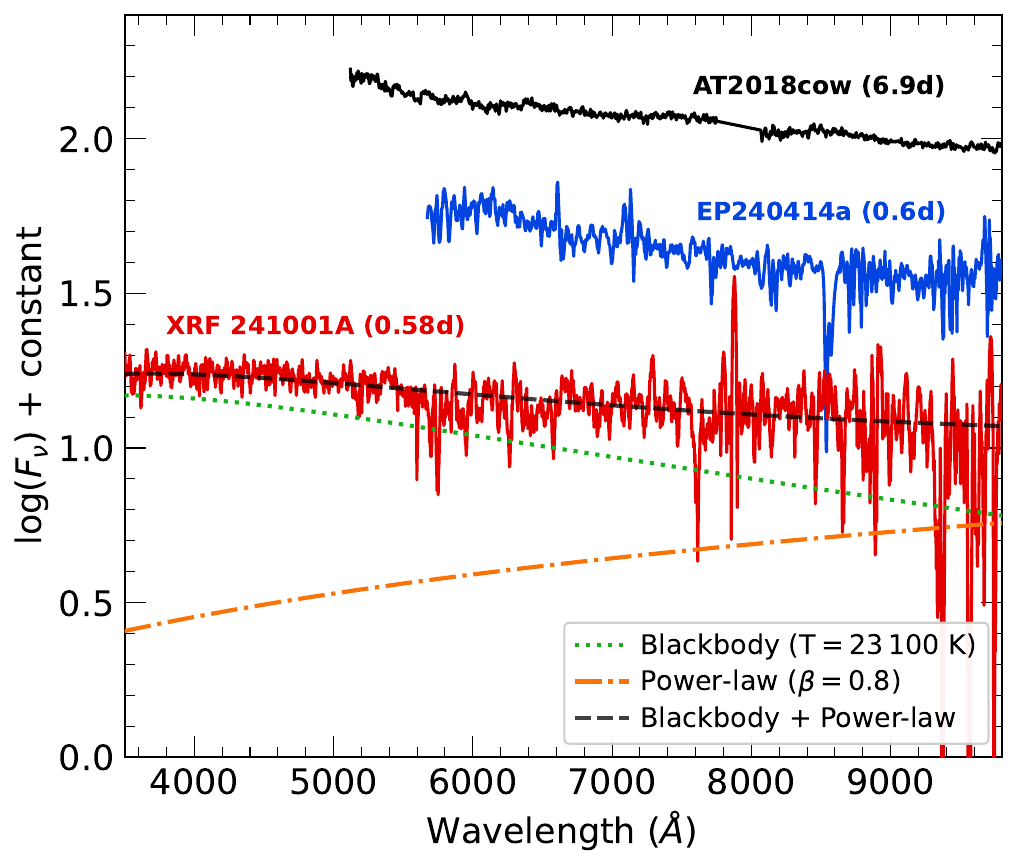}
    \caption{Ultraviolet/optical spectrum of XRF~241001A at 0.58~days (red) compared to EP240414a at 0.6~days (blue, \citealt{vandalen2025a}), AT2018cow at 6.9~days (black, \citealt{prentice2018a}), and the best-fit model (black dashed line) combining a blackbody (dotted blue line) and a power law (dash-dotted green line). Spectra of EP240414a and AT2018cow have been shifted to $z=0.5728$.}
    \label{fig:xshooter_comparison}
\end{figure}

\begin{figure*}[t]
\centering
    \includegraphics[width=17cm]{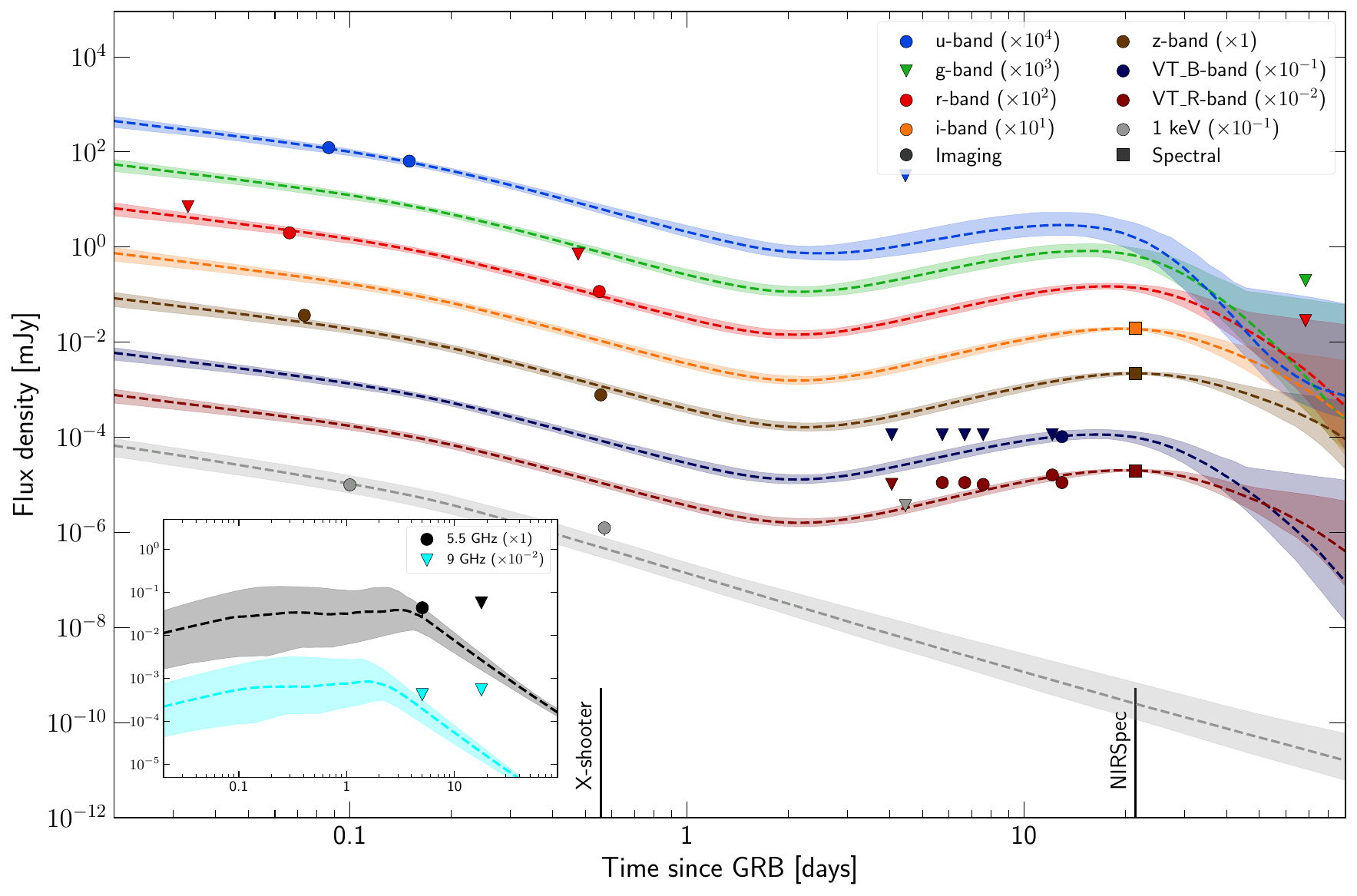}
    \caption{Multiwavelength afterglow and SN modeling of XRF~241001A and SN~2024aiiq using a top-hat jet model and an Arnett supernova model, fitted to the dataset after subtraction of the thermal component. The main panel shows the X-ray (gray) and optical light curves from the $u$ to $z$ bands. Circles indicate detections from imaging, squares represent synthetic photometry derived from spectroscopy, and triangles denote upper limits. The best-fit model for each band is shown as a dashed line, with the associated $1\sigma$ uncertainty indicated by a shaded region. The epochs of the X-shooter and JWST observations are marked by two vertical lines. The lower-left subpanel shows the radio observations and corresponding models at 5.5 GHz (black) and 9 GHz (cyan).}
    \label{fig:redback_fit_model2}
\end{figure*}
At 0.58~days after the trigger, the UV/optical part of the X-shooter spectrum reveals a strong blue excess, with only marginal signal detected in the NIR (Fig.~\ref{fig:x-shooter}). A single power-law fit yields a spectral index of $\beta = -0.4 \pm 0.1$ ($F_{\nu} \propto \nu^{-\beta}$), significantly bluer than typical values observed in GRB afterglows of $\beta \sim 0.6$ \citep[e.g.,][]{kann2010a}.

Alternatively, fitting the spectrum with a single blackbody model using {\sc Redback} \citep{sarin2024a} results in a best-fit temperature of $T = 23\,600 \pm 200$~K, capable of reproducing the strong blue emission. However, this model underestimates the flux in the visible part of the spectrum and tends to overestimate the UV part.
We then explored a two-component model consisting of a blackbody plus a power law. The best-fit model returns a temperature of $T = 23\,100 \pm 800$~K and a spectral index of $\beta = 0.8 \pm 0.5$. It provides a better fit of the data, particularly in the UV and around $\sim$8000~\AA\ (Fig.~\ref{fig:xshooter_comparison}). The inferred $\beta$ value, although poorly constrained, is consistent with those typically observed in long GRB afterglows. Our modeling strongly suggests the presence of an additional emission, likely thermal, in addition to the synchrotron emission from a relativistic jet.

In Fig.~\ref{fig:xshooter_comparison}, we also compare the spectrum of XRF~241001A with that of AT2018cow obtained with the SuperNova Integral Field Spectrograph (SNIFS; \citealt{aldering2002a}) at 6.9~days \citep{prentice2018a} and with EP240414a observed with the Gran Telescopio Canarias (GTC) at 0.6~days \citep{vandalen2025a} shifted to a common redshift of $z = 0.5728$. 
AT2018cow is a fast blue optical transient (FBOT), a rare, rapidly evolving and luminous explosion brighter and faster than typical SNe with an origin that remains uncertain \citep{prentice2018a,ho2019a,perley2019a}.
EP240414a is an eFXT detected by \EP with a duration of $T_{90} = 155$~s, an isotropic-equivalent energy of $E_{\rm iso} = 5.3 \times 10^{49}$~erg (measured in the 0.5–4~keV band), and an extremely soft spectrum with a peak energy of $E_{\rm peak} < 1.3$~keV \citep{sun2025a}. The event occurred at $z = 0.401$ and was associated with a SN Ic broad-lined (BL).
At a later epoch (6.9~days), AT2018cow appears brighter compared to XRF~241001A but exhibits a broadly similar spectral shape. Interestingly, EP240414a shows an excellent match in terms of spectral shape and luminosity to XRF~241001A at a comparable time after the trigger, similarly exhibiting an extreme optical blue spectral signature \citep{vandalen2025a}.

Additional \EP events, such as EP250108a, EP250827b or EP250304a, have shown early blue emission consistent with a blackbody emission \citep{eyles-ferris2025a,srinivasaragavan2025b,srinivasaragavan2025c,cotter2026a}.
The presence of an analogous early blue excess in multiple soft events with comparable energetics and a low-energy peak is intriguing. Further detections will shed light on whether this feature is common in these populations and whether it might point to a connection between eFXTs, XRFs, and FBOTs.

\subsection{Afterglow modeling}
\label{ssection:agmodeling}
The presence of an early blue emission component, combined with the limited early multiwavelength coverage, raises significant challenges in robustly modeling the afterglow emission.
First, we restrict our analysis to the X-ray and optical observations obtained within the first day after the trigger and explore two scenarios. 

In scenario 1 (full dataset), we assume that the emission during this period is entirely dominated by synchrotron radiation and model the dataset reported in Table~\ref{table:photometry} accordingly. 
In scenario 2 (thermal-subtracted dataset), we assume a contamination from an additional thermal component during the X-shooter observations. At this epoch, we consider only observations in the $r$  and $z$ bands, which are expected to be less affected by the blackbody emission, and we use the modeling of the spectrum presented in Sect.~\ref{ssection:blue_emission} to subtract the estimated blackbody contribution. 
At later epochs, the thermal component observed at $T_{0}+0.6$~days is still expected to contribute, but its temporal and spectral evolution cannot be constrained with our observations. We therefore excluded the VT observations obtained at 0.81~days from the dataset, as they are likely contaminated by this component and might affect the modeling of the afterglow emission.

Both datasets are modeled under the assumption that the emission is solely produced by synchrotron radiation from the forward shock formed by the interaction of the jet with a surrounding uniform density external medium \citep{sari1999b}.
We fit the X-ray to radio photometric data using {\sc Redback} v.1.15.2 \citep{sarin2024a}, considering a tophat jet structure profile implemented in {\sc VegasAfterglow} v.2.0.1 \citep{wang2026a} viewed on-axis ($\theta_v = 0\degree$) and adding host-galaxy extinction. Alternatively, we use {\sc Afterglowpy} v.0.8.1 \citep{ryan2020a} and find consistent results. We sample the parameter space using the nested sampler {\sc pymultinest} \citep{feroz2009a,buchner2014a}, implemented in {\sc Bilby} \citep{ashton2019a}. The priors adopted are provided in Table~\ref{tab:redback_params} and are sampled with a standard Gaussian likelihood, with a modification to account for non-detections implemented in {\sc Redback}. Below, we describe the afterglow properties obtained for each scenario.

The best-fit model for scenario 1 provides an overall reasonable description of the data, except at the epoch of the X-shooter observations and for the VT measurements at 0.81~days (Fig.~\ref{fig:redback_fit_model1}). At these times, the model underestimates the flux in the blue bands and overestimates it in the red bands. The posterior values of the parameters returned are reported in Table~\ref{tab:redback_params} and Fig.~\ref{fig:redback_corner_AG}.
In scenario 2, the data can be well reproduced by our model, except for the second X-ray observation, which is underestimated (Fig.~\ref{fig:redback_fit_model2}). 
The afterglow parameters for the two models are fairly typical compared to the general GRB afterglow population \citep[e.g.,][]{chrimes2022a}. 

For the dataset with the thermal component subtracted, assumed to better describe the synchrotron emission, the values indicate a relativistic outflow with an isotropic-equivalent kinetic energy of $\log(E_0/{\rm erg}) = 52.00^{+1.00}_{-0.83}$, expanding into a moderately dense circumburst medium with $\log(n_0/\mathrm{cm^{-3}}) = 0.52^{+1.16}_{-1.38}$ ($n_0 \sim 3.3 \,\mathrm{cm^{-3}}$). The inferred half jet core opening angle is $\theta_c = 4.0^{+2.9}_{-1.7}$~deg (Table~\ref{tab:redback_params} and Fig.~\ref{fig:redback_corner_AG}). 
In both models, allowing the jet viewing angle $\theta_v$ to vary does not significantly improve the fit, and the results remain consistent with an on-axis jet ($\theta_v < \theta_c$). For scenario~2, the Bayesian evidence decreases from $\log Z = 38.09$ to $\log Z = 33.83$ when $\theta_v$ is allowed to vary.
Overall, our results disfavor an off-axis interpretation and suggest that the low luminosity observed is intrinsic rather than dominated by geometric effects. 
Our modeling also provides a constraint on the extinction along the line of sight within the host galaxy. The two models return a $A_V^{\rm GRB} < 0.05$ mag, indicating negligible host dust attenuation. We therefore adopted $A_V^{\rm GRB} = 0$ mag for the rest of the analysis.

In our modeling, we adopt an external medium of uniform density that can provide a reasonable fit of the data. However, the association of the burst with a SN Ic-BL (Sect.~\ref{ssection:SupernovaAnalysis}) indicates a massive star progenitor that could produce a wind-like environment. Fitting the data with a wind-like profile does not significantly improve the fit and yields similar parameter constraints, with a wind density $A_\ast = 0.30^{+0.11}_{-0.08}$ (full dataset) and $0.03^{+0.07}_{-0.02}$ (thermal-subtracted dataset). 
In scenario~2, the wind-like scenario provides a larger Bayesian evidence ($\log Z = 45.74$) than the ISM case ($\log Z = 38.09$), while yielding comparable parameters such as $\log(E_0/{\rm erg}) \sim 52.36$ (in contrast to $52.00$) and $\theta_c \sim 5.2^\circ$ (compared to $\sim$$4.0^\circ$). We choose the uniform density medium as our reference while noting that a wind-like medium remains a possible alternative.

Recently, some events have shown a possible preference for structured jet \citep[e.g.,][]{oconnor2023a,li2025a,srinivasaragavan2025a}. For the thermal subtracted dataset, we additionally consider a jet with a Gaussian structure implemented in {\sc VegasAfterglow}. We find consistent results with the top-hat jet model (Figs.~\ref{fig:redback_fit_model_gauss} and \ref{fig:redback_corner_AG_tophat_gaussian}), with Bayesian evidences nearly identical for the two models ($\log Z = 38.09$ for the top-hat jet and $\log Z = 38.07$ for the Gaussian jet). Similarly, allowing $\theta_v$ to vary does not improve the fit quality ($\log Z = 33.54$), and we consistently obtain $\theta_v < \theta_c$.

Finally, we also investigate the possible effect of the synchrotron self-Compton (SSC) emission by enabling this component in {\sc VegasAfterglow}, but we find no significant improvement in the fit, nor substantial changes in the inferred parameters.

\subsection{X-ray to NIR afterglow SED fitting}
\label{ssection:SEDfit}
\begin{figure}[t]
    \centering
    \includegraphics[width=\hsize]{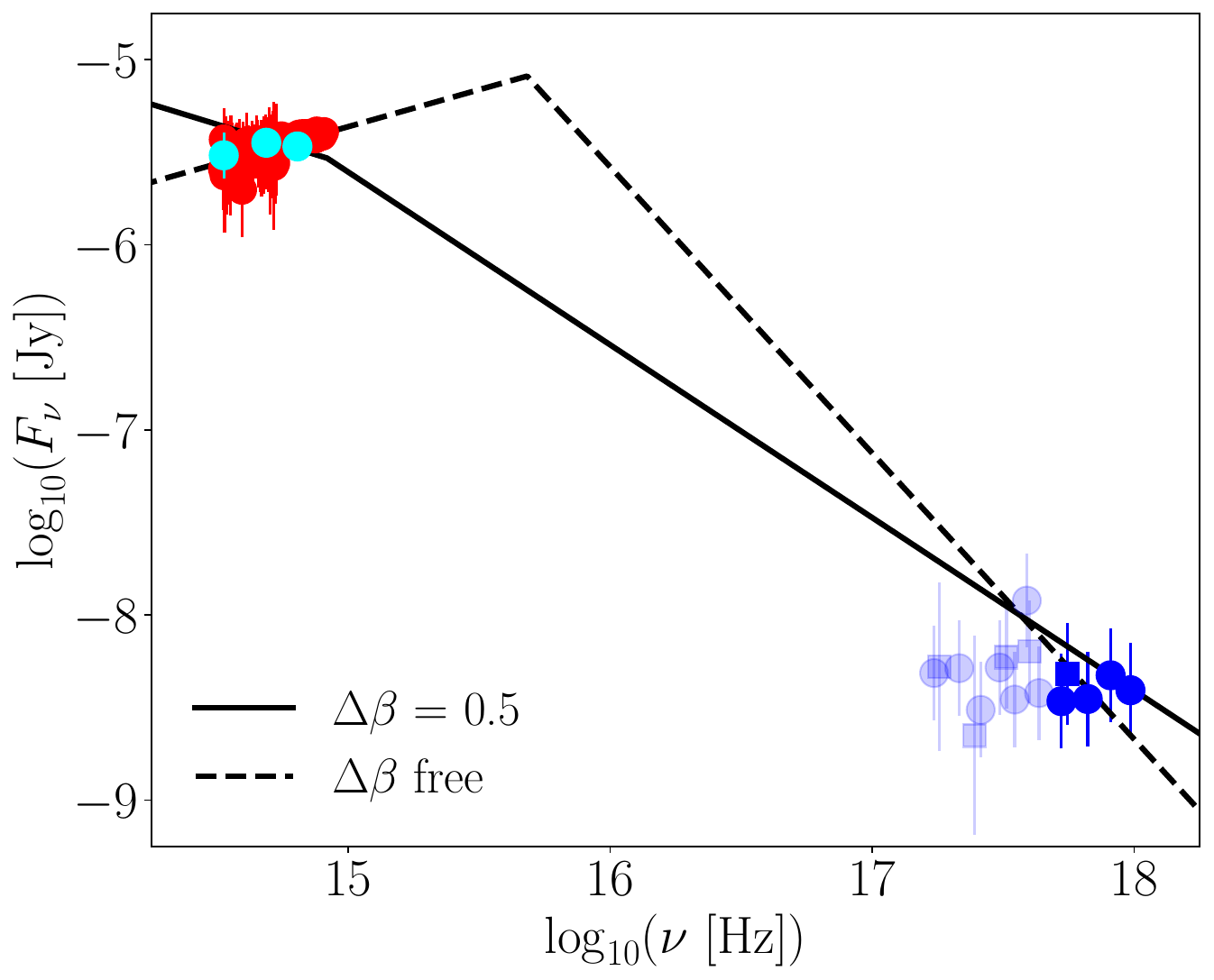}
    \caption{X-ray–to–NIR SED modeling of XRF~241001A at 13.4~hours after the trigger. The photometry-calibrated X-shooter spectrum is shown in red circles, with photometric detections indicated by cyan circles. The \EP/FXT and extrapolated \Swift/XRT spectra are shown as blue squares and circles, for which the shaded data points ($<$2~keV) were excluded in the fit to limit the effect of photoelectric absorption. The black dashed line shows the best-fit model obtained assuming a free $\Delta\beta$, while the black solid line corresponds to a model with a fixed spectral change of $\Delta\beta =\beta_\mathrm{X}-\beta_\mathrm{O}=0.5$.}
    \label{fig:xshooterSED}
\end{figure}

\begin{figure*}[t]
    \begin{subfigure}[b]{0.49\hsize}
        \centering
        \includegraphics[width=\hsize]{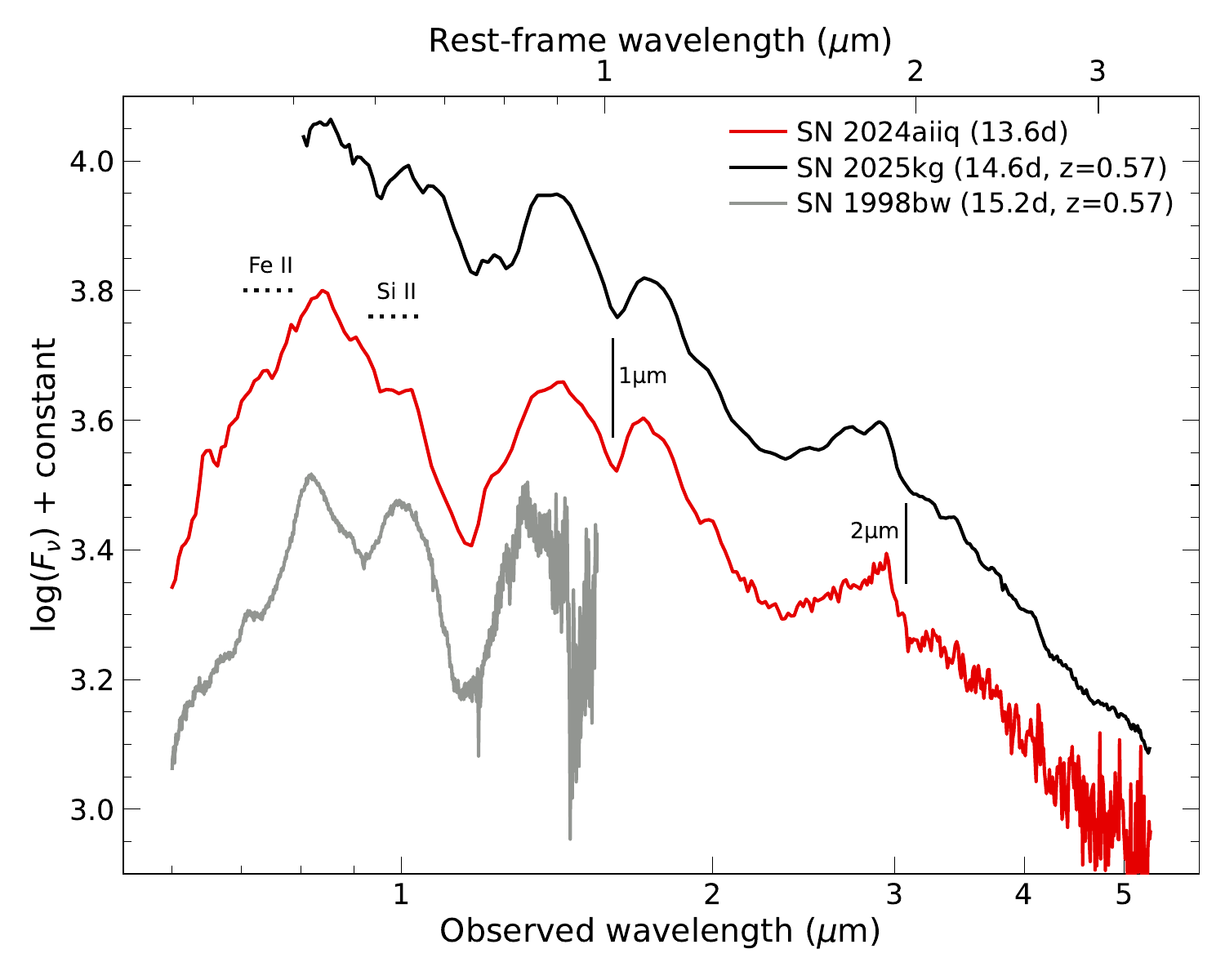}
    \end{subfigure}
    \hfill
    \begin{subfigure}[b]{0.49\hsize}
        \centering
        \includegraphics[width=\hsize]{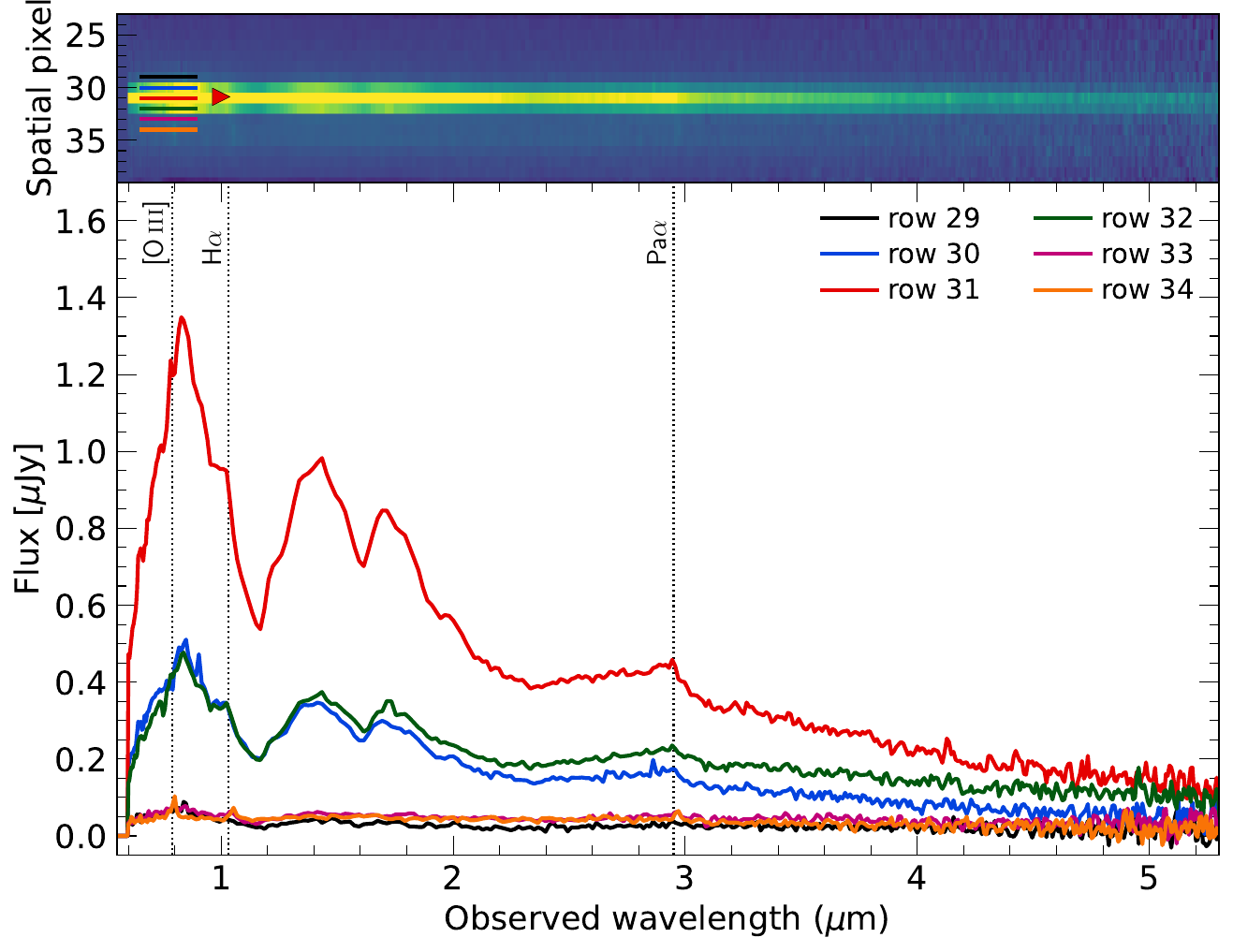}
    \end{subfigure}
    \caption{
        \textit{Left panel:} Spectrum from \JWST/NIRSpec of XRF~241001A/SN~2024aiiq (red) compared with the archetypal GRB/SN Ic-BL GRB~980425/SN~1998bw (gray, \citealt{patat2001a}) and EP~250108a/SN~2025kg (black, \citealt{rastinejad2025a}) observed at a similar rest-frame epoch. The Fe~II~$\lambda5169$ and Si~II~$\lambda6355$ features are marked with dashed horizontal lines, and the broad features near 1 and 2~$\mu$m are indicated with vertical lines.
        \textit{Right panel:} Two-dimensional (top) and extracted one-dimensional (bottom) \textit{JWST}/NIRSpec spectra of XRF~241001A/SN~2024aiiq. The color-coded lines mark the spatial rows used for the 1D extraction, the red arrow indicates the best afterglow position, and dashed vertical lines show the expected positions of host strong emission lines.}
    \label{fig:jwst_spec}
\end{figure*}

The shape of the X-ray-to-NIR afterglow continuum is determined by using the \EP/FXT, \Swift/XRT and VLT/X-shooter data to create an SED. The SED epoch is chosen at $\sim$13.4~hours after the trigger, matching the quasi-simultaneous X-shooter and \EP/FXT observations. To benefit from the early XRT observations with higher S/N values, we scaled the XRT spectrum to the same epoch using a normalization factor derived from the X-ray light curve model, assuming a single temporal power law decay.
To further increase the S/N, the \EP/FXT and \Swift/XRT spectral files are grouped to at least 3 counts per energy channel in our analysis. We smoothed and increased the S/N of the X-shooter spectrum by grouping 300 data points per wavelength bin. The low S/N bins, typically above 9000 \AA{} (observer frame), are excluded during the fit, and the resulting spectrum is corrected for Galactic extinction ($A_V^{\rm Gal} = 0.03$ mag, \citealt{schlafly2011a}).

Considering the faintness of the X-ray afterglow observed (Fig.~\ref{fig:AG_comparison}), we choose to perform our analysis using the X-ray energy ranges where the effect of photoelectric absorption is minimal (i.e., $>$2 keV). We also assume no host-galaxy dust extinction ($A_V^{\rm GRB} = 0$) along the line of sight. This choice is motivated by the results of the afterglow modeling (Sect.~\ref{ssection:agmodeling}), indicating a small dust extinction from the host galaxy ($A_\mathrm{V}< 0.1$ mag)
To estimate the spectral intrinsic parameters of the X-ray to NIR afterglow, we use a Markov chain Monte Carlo algorithm and the fitting method described in \cite{rakotondrainibe2024a}. We model the continuum using a single power-law model and a broken power-law model. The broken power-law model is constructed assuming that the cooling frequency, $\nu_\mathrm{c}$, is located between the UV-to-X-ray energy ranges and that the spectral break is fixed at $\Delta\beta =\beta_\mathrm{X}-\beta_\mathrm{O} = 0.5$ \citep{sari1998a}. Alternatively, we also let $\Delta\beta$ free, allowing both slopes to vary independently. A fit with a single power-law cannot reproduce the observed X-ray-to-NIR data and is rejected. 

In Fig.~\ref{fig:xshooterSED}, we show the results that include the fixed and free break best-fit models. Only the model with the spectral break allowed to vary freely can reasonably fit the data with $\beta_\mathrm{X} = 1.60^{+0.36}_{-0.17}$ and $\beta_\mathrm{O} =- 0.40^{+0.16}_{-0.13}$. Fitting the SED with the fixed spectral break poorly fits the observed data by overestimating and underestimating the UV-NIR and X-ray data, respectively \citep[see][]{zafar2011a}. This model gives $\beta_\mathrm{X} = 0.92^{+0.02}_{-0.02}$, with $\beta_\mathrm{O} = \beta_\mathrm{X}- 0. 5 = 0.42$.
The tension observed with the cooling break scenario further supports the presence of an additional early blue component superimposed on the afterglow emission, as discussed in Sect.~\ref{ssection:blue_emission}.

\subsection{Supernova analysis}
\label{ssection:SupernovaAnalysis}
The late-time re-brightening observed around 10 days by the SVOM/VT is compatible with the expected rise from an associated SN, hereafter SN~2024aiiq. We model the observations using a one-zone radioactive decay model following \cite{arnett1980a,arnett1982a}. Given the nonnegligible uncertainties in our afterglow modeling, we fit the SN component jointly with both scenarios described above. For each model, we adopted the posterior distributions from the best-fit model and restricted the parameters to the 16th–84th percentile interval of their solutions.

The best-fit models are shown in Fig.~\ref{fig:redback_fit_model2} and Fig.~\ref{fig:redback_fit_model1} for the joint fits corresponding to scenario 2 and scenario 1, respectively. Both models successfully reproduce the SN bump, although they tend to underestimate the first VT observation obtained during the rising phase. 
We find that the two scenarios provide consistent posterior distributions (Table~\ref{tab:redback_params}, Fig.~\ref{fig:redback_corner_SN}), indicating a negligible effect of the afterglow modeling on the inferred SN parameters.

For the joint fit with scenario 2 (thermal component subtracted), SN~2024aiiq is characterized by an ejecta mass of $M_{\rm ej} = 2.59^{+0.91}_{-0.60}\,M_\odot$ and a nickel mass fraction of $f_{\rm Ni} = 0.20^{+0.05}_{-0.04}$, corresponding to the production of $M_{\rm Ni} = 0.52^{+0.05}_{-0.03}\,M_\odot$.
Compared to Type Ic-BL SNe associated with GRBs, which typically synthesize $\sim0.4 \pm 0.2\,M_\odot$ of $^{56}$Ni \citep[e.g.,][]{cano2017a,lu2018a}, and ordinary Type Ic-BL SNe (without GRBs) producing $\sim0.3 \pm 0.2\,M_\odot$ \citep[e.g.,][]{taddia2019a,srinivasaragavan2024b}, SN~2024aiiq lies within the range of GRB/SNe, while occupying the high end of the ordinary Ic-BL distribution.
At the moment, the comparison of SN~2024aiiq with the XRF/SNe population remains limited by the small sample size and the heterogeneity of events, see \citet{srinivasaragavan2025c} and references therein for a recent discussion on XRF/SNe.

We measured an ejecta velocity of $v_{\rm ej} = (2.32^{+0.40}_{-0.32})\times10^4\,\mathrm{km\,s^{-1}}$. Assuming a spherical thin-shell explosion in which all ejecta move at the same velocity ($E_{\rm SN} = \tfrac{1}{2} M_{\rm ej} v_{\rm ej}^2$), we derived a kinetic energy of $E_{\rm SN} = (1.39^{+0.68}_{-0.50}) \times 10^{52}\,\mathrm{erg}$. Adopting a homologous uniform density model with a linear velocity profile \citep[e.g.,][]{wheeler2015a,lyman2016a} gives $E_{\rm SN} = (0.83^{+0.41}_{-0.30}) \times 10^{52}\,\mathrm{erg}$, using the relation $E_{\rm SN} = \tfrac{3}{10} M_{\rm ej} v_{\rm ej}^2$. Both estimates are consistent with the range observed for GRB/SNe, although toward the lower end of the distribution \citep{lu2018a, finneran2025a}.

We caution that the temporal coverage of SN~2024aiiq is limited, which might affect the inferred parameters. This is partially represented by the uncertainty in the marginalized posterior, but it misses the systematics associated with the model. Our approach of fitting the afterglow and SN components simultaneously differs from the more commonly adopted method in which the two components are modeled independently. While, as noted above, this choice appears to have a negligible impact on the results, it may still introduce systematic differences that can only be robustly quantified with a larger sample analyzed in a homogeneous manner, which is beyond the scope of this work.

The \textit{JWST}/NIRSpec near-infrared spectrum shows a clear detection of a continuum with broad absorption features characteristic of a SN component. To identify and classify the SN, we performed spectral matching using \textsc{SNID} \citep{blondin2007a} with the Super-SNID template set \citep{magill2025a}, which incorporates Type Ic-BL templates from \cite{modjaz2016a}, commonly observed in association with long-duration GRBs \citep{cano2017a}. 
The best spectral match corresponds to the archetypal GRB/SN Ic-BL SN~1998bw \citep{patat2001a} at a phase of $\pm1$ days relative to peak brightness (left panel of Fig.~\ref{fig:jwst_spec}). 
The \textsc{SNID} templates are restricted to wavelengths below 1~$\mu$m. Although extended templates up to 2~$\mu$m can be provided, the comparison of the \textit{JWST} spectrum beyond this range remains limited by the small number of templates available.

To extend the comparison into the near-infrared,  we considered the \textit{JWST} spectrum of SN~2025kg, associated with EP250108a and classified as a SN Ic-BL \citep{rastinejad2025a}. We found excellent agreement in both the continuum shape and the broad spectral features, particularly in the observed range of 1.5 to 5~$\mu$m.
Our spectral comparisons indicate that a SN Type Ic-BL provides the best match to the data, with a phase consistent with observations near maximum light. The identification of a SN associated with XRF~241001A firmly establishes it as a collapsar event arising from the core collapse of a massive star.

\begin{figure*}[t]
    \begin{subfigure}[b]{0.49\hsize}
        \centering
        \includegraphics[width=\hsize]{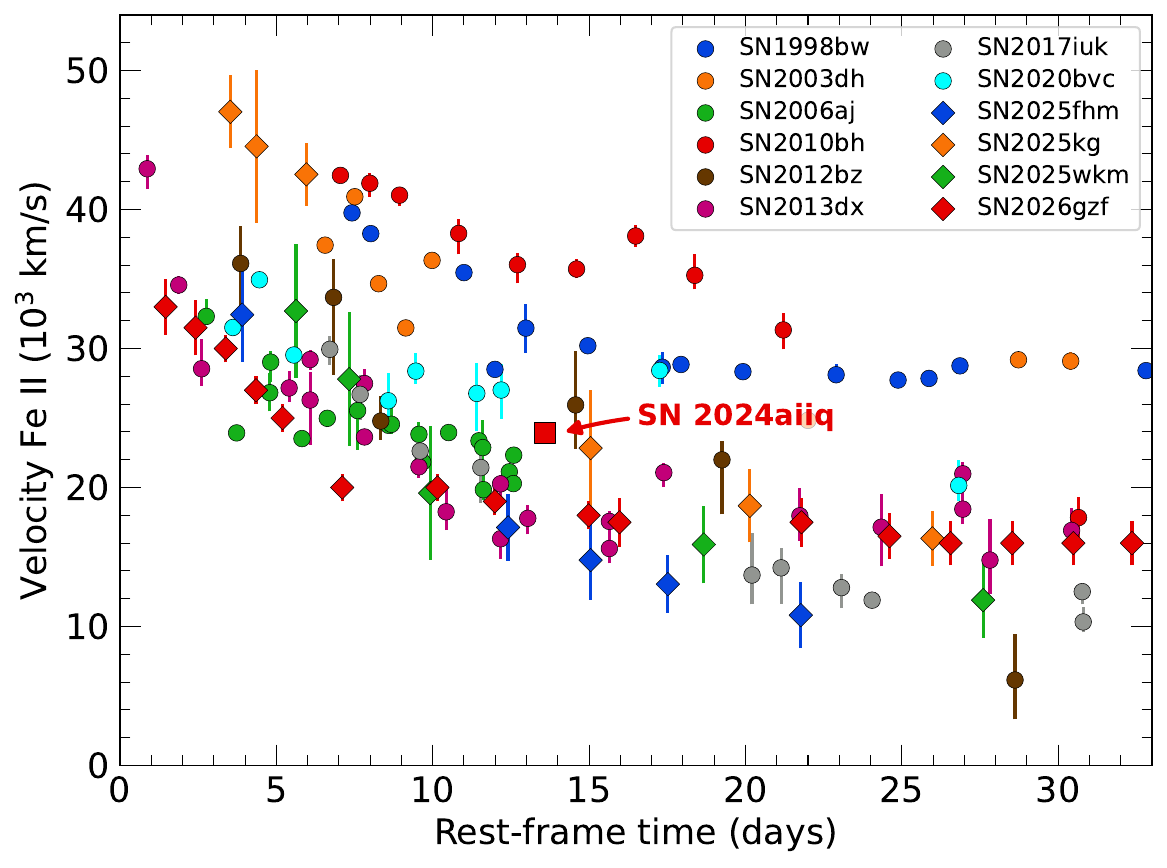}
    \end{subfigure}
    \hfill
    \begin{subfigure}[b]{0.49\hsize}
        \centering
        \includegraphics[width=\hsize]{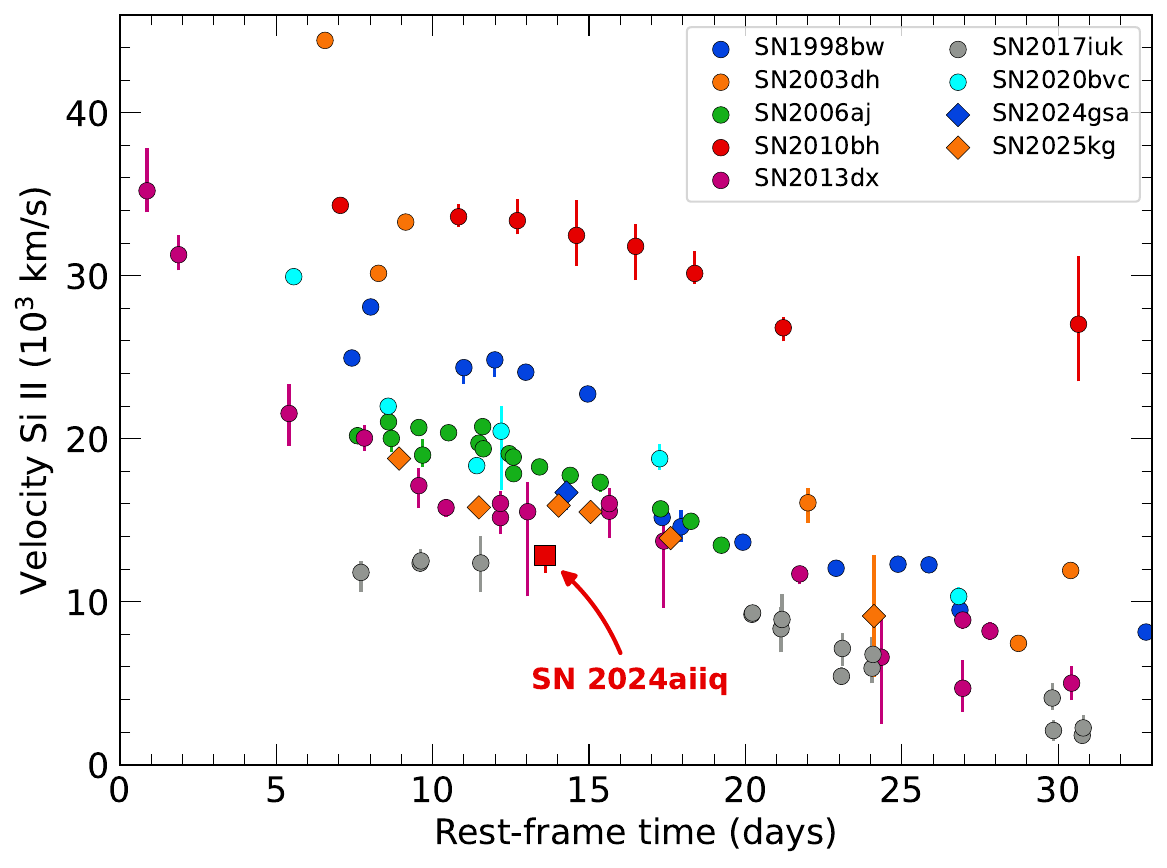}
    \end{subfigure}
    \caption{Temporal evolution of expansion velocities in GRB/SNe and eFXT/SNe.
    \textit{Left panel:} Fe~II~5169~\angstrom\ velocity as a function of rest-frame time for a sample of GRB/SNe (circles) from \cite{finneran2025b} and eFXT/SNe (diamonds) from \cite{rastinejad2025a,sun2025a,srinivasaragavan2025b,srinivasaragavan2025c,cotter2026a,martin-carrillo2026a}, with XRF~241001A/ SN~2024aiiq highlighted as a red square. 
    \textit{Right panel:} Same as the left panel but for the Si~II~6355~\angstrom\ line.}
    \label{fig:SN_velocity}
\end{figure*}
The JWST spectrum of SN~2024aiiq exhibits several broad absorption features characteristic of SNe Ic-BL. They are believed to result from the rapid expansion of the outer ejecta layers, with typical velocities of $\sim$15,000–20,000~km~s$^{-1}$ around maximum light \citep{modjaz2016a}. The broadening of the line leads to the blending of adjacent lines, which challenges the precise determination of the expansion velocities.
The Fe~II and Si~II lines are among the most prominent absorption features and are commonly used as tracers of the photospheric expansion velocity in SNe Ic-BL. 
The Fe~II feature is known to be blended with at least three transitions at rest-frame wavelengths of 4924~\AA, 5018~\AA, and 5169~\AA\ \citep{modjaz2016a}. In the case of Si~II, the canonical rest-frame wavelength is 6355~\AA. However, blending with Na~I may affect the line profile, as noted by \citet{modjaz2016a}.

We note the possible presence of narrow emission peaks at the expected wavelengths of several nebular lines, such as \OIIIab, which are also visible in the X-shooter spectrum (Fig.~\ref{fig:x-shooter}). This suggests that the spectrum might be contaminated by strong emission lines of the host galaxy (see Sect.~\ref{ssect:host_gal}).
Similarly, contamination from the host continuum is difficult to assess. However, the observed SN magnitude of $r = 23.24 \pm 0.05$ AB mag, compared to the constraint given by archival images ($r > 25.4$ AB mag, see also Sect.~\ref{ssect:host_gal}), suggests a negligible contribution at optical wavelengths. This is supported by the row-by-row decomposition (right panel of Fig.~\ref{fig:jwst_spec}), where adjacent rows to the SN dominated one (row 31) exhibit similar profiles.

For consistency and direct comparison with previous work, we adopted the same method as \citet{finneran2025b} to measure the expansion velocities from the blueshifted absorption minima of the Fe~II 5169~\AA\ and Si~II 6355~\AA\ features. For each line, we selected a wavelength interval bracketing the absorption trough and fitted a cubic spline to the observed spectrum within this region. The wavelength of the absorption minimum was determined as the minimum of the spline evaluated on a finely sampled grid. The corresponding expansion velocity was computed using the relativistic Doppler formula. Uncertainties on the velocities were estimated via a Monte Carlo approach. We generated $N=10,000$ realizations of the spectrum by perturbing the flux at each wavelength according to its measured uncertainty and repeated the spline fitting and minimum determination for each realization. The median of the resulting velocity distribution was adopted as the best-fit value, with the 16th and 84th percentiles providing the lower and upper uncertainties, respectively.

From the Fe~II~5169~\angstrom\ absorption minimum, we measure an expansion velocity of $v_{\mathrm{Fe\,II}} = 23{,}949^{+360}_{-467}\,\mathrm{km\,s^{-1}}$, while the Si~II~6355\,\angstrom\ feature yields 
$v_{\mathrm{Si\,II}} = 12{,}837^{+467}_{-1{,}066}\,\mathrm{km\,s^{-1}}$.
In Fig.~\ref{fig:SN_velocity}, we compare the single-epoch measurements of SN~2024aiiq with the temporal evolution of the Fe~II~5169~\angstrom\ (left panel) and Si~II~6355~\angstrom\ (right panel) expansion velocities for a sample of GRB/SNe measured uniformly by \cite{finneran2025b}. We also include recently published eFXT/SNe detected by \EP from \cite{rastinejad2025a,sun2025a,srinivasaragavan2025b,srinivasaragavan2025c} for comparison, although we caution that the use of different velocity measurement methods may introduce systematics.
The Fe~II velocity of SN~2024aiiq is consistent with the range observed in GRB/SNe near maximum light, while the Si~II velocity appears toward the lower end of the distribution.
Overall, the measured velocities of SN~2024aiiq are consistent with those of typical GRB/SNe. 
The Fe~II velocity is higher than that inferred from Si~II at comparable epochs ($\sim$$11\,000 \,\mathrm{km\,s^{-1}}$), as commonly observed in broad-lined events, possibly suggesting that these transitions trace material at different layers within the ejecta \citep{modjaz2016a,finneran2025b}.

Similarly to Sect.~\ref{ssection:optical_afterglow_analysis}, we derived a synthetic $V$ band absolute magnitude from the \textit{JWST} spectrum. We obtained $M_V = -19.15 \pm 0.05$ mag at the epoch of observation. When compared to the distribution of GRB/SNe compiled in the GRBSN Webtool \citep{finneran2025a}, this value lies within the typical range observed near maximum light. The inferred luminosity is consistent with previously reported GRB/SNe and does not indicate an unusually faint or exceptionally luminous event. The low luminosity of the prompt and afterglow emission, in contrast with the luminosity of SN~2024aiiq, which is typical of that observed for SNe associated with GRBs, further supports the lack of a correlation between GRB properties and their associated SNe \citep[e.g.,][]{tanvir2010a,melandri2014a,michalowski2018b,srinivasaragavan2024a}.

Notably, SN~2024aiiq also exhibits broad absorption features around $\sim$1~$\mu$m and $\sim$2~$\mu$m (Fig.~\ref{fig:jwst_spec}) that are strikingly similar to those observed in EP250108a/SN~2025kg \citep{rastinejad2025a} and show similarities to the NIR spectrum of SN~1998bw \citep{patat2001a}. As discussed for SN~2025kg, the feature near 1~$\mu$m may arise from a blend of multiple species, including He I 1.0830~$\mu$m, C I 1.0693~$\mu$m, and Mg II 1.0927~$\mu$m, while the shallower absorption near 2~$\mu$m could be associated with He I 2.0581~$\mu$m. The broader morphology of the 1~$\mu$m feature compared to the 2~$\mu$m feature is consistent with a blended origin rather than a single transition. The close resemblance between the NIR spectra of SN~2024aiiq and SN~2025kg suggests that similar physical processes are involved, potentially including the presence of a small amount of helium in the ejecta. However, the interpretation remains limited by line blending and the still sparse NIR spectroscopic coverage of GRB/SNe. Further investigation with a larger sample of GRB/SNe observed with NIR coverage will be essential to assessing the prevalence and origin of these features.

\subsection{Host galaxy}
\label{ssect:host_gal}
Our VLT/X-shooter spectrum reveals the presence of multiple emission lines from the underlying host galaxy. We clearly detected the \OII doublet, \OIIIa, and \OIIIb. A faint \hbeta line was also detected, though it is contaminated by telluric lines. We corrected the line for telluric absorption using the sky model provided by \textsc{SKYCALC}\footnote{\url{https://www.eso.org/observing/etc/bin/gen/form?INS.MODE=swspectr+INS.NAME=SKYCALC}} v2.0.9 \citep{noll2012a,jones2013a} at the target coordinates and observing time. The \halpha line falls into a poor S/N of the NIR spectrum, and only a hint of the line was detected (<$3\sigma$).

We corrected the spectrum for Milky Way dust extinction \citep[$A_V^{\rm Gal} = 0.03$~mag,][]{schlafly2011a} and normalized it using the scaling factor obtained from photometry observations of the X-shooter acquisition camera (Sect.~\ref{ssect:x-shooter}). The blue afterglow observed by X-shooter and the low $A^{\mathrm{GRB}}_V \sim 0$~mag along the GRB line of sight returned by the afterglow modeling (Sect.~\ref{ssection:agmodeling}) suggest negligible dust attenuation of the host, and no correction was applied. The fluxes of each line were then measured by a Gaussian fit, and the measurements are reported in Table~\ref{tab:host_em_line_flux}. Alternatively, we also integrated the continuum-subtracted flux over the region of the line and confirmed the consistency between the two measurements. 

We used the upper limit on the \halpha flux to determine an upper limit on the star formation rate (SFR) of the host galaxy using the relation of \citet{kennicutt1998a}, scaled to the initial mass function of \citet{chabrier2003a}, which yields ${\rm SFR} < 0.3~\Msun\,\rm yr^{-1}$.
Assuming a case~B recombination \citep{osterbrock1989a} and no dust correction, we also used the \hbeta flux to derive a lower limit ${\rm SFR} > 0.1~\Msun\,\rm yr^{-1}$.
Using the relation between the \OIIIb to \OII ratio and the gas-phase metallicity of \citet{maiolino2008a}, we find that the host has a low metallicity of $12+\log(\rm O/H) = 8.0^{+0.2}_{-0.3}$, which is also confirmed by the lower limit on the \OIIIb to \NIIb ratio ($12+\log(\rm O/H) < 8.8$).
These properties are similar to those of the hosts of long GRBs derived using the same calibrators \citep{kruhler2015a,vergani2017a,palmerio2019a}, as expected considering the collapsar origin of this event.

No significant differences have been observed between XRF and long GRB host galaxies \citep{bi2018a}. Long GRBs are known to occur preferentially in the bright, inner regions of their host galaxies, typically associated with intense star formation and exhibiting relatively small projected offsets from the host centers \citep{fruchter2006a,lyman2017a,schneider2022a}. Given the collapsar origin of this event, similar to the long GRB population, we expect it to be located close to a star-forming region of its host galaxy, with a correspondingly small projected offset.
In archival DESI Legacy Survey images \citep{dey2019a}, no host galaxy is visible at the burst position down to 3$\sigma$ upper limits of $g>25.4$, $r>25.1$, $i>24.3$, and $z>23.8$ AB magnitudes, as well as in our deep Magellan/LDSS images ($g>25.7$ and $r>25.4$ at 3$\sigma$). Compared to the host galaxies of long GRB in the TOUGH sample \citep{hjorth2012a}, the host of XRF~241001A lies at the faint end of the distribution for its redshift.
Adopting the upper limit in the $i$ band as the closest proxy to the rest-frame $B$ at $z=0.5728$, we derive an absolute magnitude limit of $M_B \gtrsim -17.9$. Considering the rest-frame $B$ band luminosity function at $z\sim0.5$ from \cite{faber2007a}, this corresponds to $L_B \lesssim 0.05\,L_B^*$, placing the host in the faint, sub-$L^*$ regime, as expected from its low metallicity \citep[e.g.,][]{vergani2015a,perley2016c}.

We investigated the possible presence of strong emission lines from the host galaxy in the JWST/NIRSpec spectrum. In the right panel of Fig.~\ref{fig:jwst_spec}, we mark the wavelengths corresponding to \OIII, \halpha, and \paalpha at $z_{\textrm{em}} = 0.5729$. 
Near the \OIII doublet (blended at the NIRSpec resolution), we detect a marginal rising feature spanning multiple spatial rows 31-32, which may be associated with emission from the host galaxy in addition to the underlying SN contribution. A faint feature is also visible at the expected \paalpha wavelength, where contamination from the SN is expected to be weaker, extending over adjacent rows 30-32 and potentially consistent with \paalpha emission from the host galaxy.

Finally, in spatial row 34, where the contribution from the SN emission is negligible, we detect several nebular lines systematically shifted to longer wavelengths relative to their expected positions at $z_{\textrm{em}}$. We identify these features as \hbeta, \OIII doublet, \halpha, and \paalpha, corresponding to a mean redshift of $z_{\textrm{mean}} = 0.5951 \pm 0.0069$.
Given the NIRSpec spatial pixel scale of 0.1$\arcsec{}$/pixel, corresponding to 0.67 kpc/pix at $z_{\textrm{em}}$, the 3-pixel separation between these emission lines and the SN position (row 31) implies a projected separation of $\sim$2~kpc and suggests an origin physically distinct from the GRB/SN site. 
The inferred velocity offset of $\sim$4000~km\,s$^{-1}$The best spectral match strongly disfavors an origin within the same host galaxy and instead suggests a separate object. The absence of deep, high-resolution imaging of the field prevents us from establishing the nature of this configuration, in particular, whether the offset emission is associated with a companion system or results from a chance projection along the line of sight.

\section{Discussion}
\label{sec:Discussion}

The event XRF~241001A shows a prompt emission dominated by photons below $\sim$20~keV (Fig.~\ref{fig:eclairs_lc}), which can be fitted by a blackbody model, suggesting a possible thermal origin for the emission. Such a scenario could be interpreted in the context of relativistic shock breakout. However, the low S/N of the prompt signal does not allow us to discriminate between thermal and nonthermal models, commonly observed for long GRBs powered by relativistic jets. In addition, the apparent variability in the prompt emission light curve tends to disfavor a relativistic shock breakout interpretation, which is typically expected to produce a smoother signal.

To test whether the observed prompt emission might be consistent with a relativistic shock breakout origin, we apply the closure relation reported by \cite{nakar2012a}, linking the observed duration, total radiated energy, and breakout temperature of a relativistic shock breakout flare.
In this scenario, soft and energetic events are expected to have correspondingly long durations. For comparison, XRFs~060218 and 100316D, with $E_{\rm iso} \sim 5 \times 10^{49}$~erg and $E_{\rm peak} \sim 40$~keV, yield predicted durations of the order of $10^{3}$~s, in agreement with their observed long-lasting emission \citep{kaneko2007a,starling2011a,nakar2012a}. Applying the same reasoning to XRF~241001A, with $E_{\rm iso} = 7.15 \times 10^{49}$~erg and $E_{\rm peak} = 6.7$~keV, the expected breakout duration would be many orders of magnitude longer than the observed $T_{90} = 3.14$~s, making it inconsistent with a relativistic shock breakout.

Alternatively, assuming the prompt emission is thermal, we explore a possible connection between the early and later blackbody components. The prompt spectrum is consistent with a thermal component characterized by $kT_1 = 1.72$~keV, corresponding to a temperature $T_1 \approx 2.0\times10^{7}$~K and a radius $R_1 \approx 2.2\times10^{11}$~cm. At $T_0+0.58$~days, the UV/optical spectrum reveals a thermal component with $T_2 \approx 2.3\times10^{4}$~K and $R_2 \approx 5.4\times10^{14}$~cm. Although these components are detected in very different spectral regimes, their inferred properties are not inconsistent with those expected from an expanding and cooling thermal emitter. In particular, the inferred radius increases by nearly three orders of magnitude over this interval, corresponding to an average expansion velocity of $v \sim 0.4c$. Such mildly relativistic expansion is comparable to the early cooling phases suggested in some soft X-ray events, such as EP250108a, where rapidly expanding blackbody emission might originate from a shocked cocoon emerging from a trapped jet \citep[e.g.,][]{eyles-ferris2025a,zhu2025a}. In this case, the two thermal components may represent different phases of an expanding outflow, where the prompt emission could arise from the jet photosphere or from shocked cocoon material \citep[e.g.,][]{nakar2015a,nakar2017a,zheng2026a}, while the later thermal emission may trace the subsequent expansion and cooling of this material. However, we caution that the available data do not allow us to further investigate whether these signatures arise from the same evolving component.

In contrast, our multiwavelength modeling of the afterglow emission suggests the formation of a relativistic jet observed on-axis and interacting with a moderately dense circumburst medium. 
In this context, the prompt emission can be naturally interpreted as arising from internal dissipation within the jet, favoring nonthermal models typically associated with relativistic jet.
The detection of an associated SN Type Ic-BL firmly establishes its origin from the core collapse of a massive star. The SN~2024aiiq properties are consistent with those typically observed in long GRBs, exhibiting harder and more energetic prompt emissions.

Combining the prompt and afterglow analyzes, we derive a GRB efficiency \citep[e.g.,][]{lloyd-ronning2004a} of $\zeta = 0.02^{+0.02}_{-0.01}$ for the full dataset and $\zeta = 0.01^{+0.04}_{-0.01}$ for the dataset with the thermal component subtracted. Both models suggest a relatively low gamma-ray efficiency below $\sim$2\%. We note, however, that these estimates remain uncertain given the limited dataset and the assumptions involved in our modeling.
Similar low efficiencies have been reported for soft X-ray transients such as EP240414a \citep{sun2025a}, although in that case, the afterglow evolution includes a rebrightening. For these soft events, the low apparent efficiency might be interpreted in terms of off-axis viewing from a structured jets with a narrow jet core \citep{ren2026a}, from a jet-cocoon viewed off-axis \citep{zheng2025a} or possibly by a jet surrounded by a dense circumstellar material \citep{hamidani2025a}. By comparison, XRF~241001A does not show evidence of a rebrightening, and the afterglow modeling favors a jet viewed on axis, indicating that the low gamma-ray efficiency is likely intrinsic to the outflow rather than driven by geometric effects.

Our results suggest that XRF~241001A was produced by a massive star and was powered by a weak relativistic jet observed on-axis with low gamma-ray efficiency production. In this scenario, this event would belong to the soft, low-luminosity tail of the long GRB distribution \citep[e.g.,][]{barraud2005a} and could potentially be associated with a baryon-loaded GRB with a low Lorentz factor, often referred to as a “dirty fireball” \citep{dermer1999b}. 
While this scenario has been proposed for the eFXT EP241113a \citep{dai2026a}, the dataset of XRF~241001A is insufficient to robustly establish a dirty fireball origin.
Future early- and late-time multiwavelength follow-up, including sensitive radio observations, is essential to assess whether some XRFs are genuinely associated with dirty fireball scenarios.
In particular, broadband afterglow evolution could constrain the kinetic energy and initial Lorentz factor of the jet through the afterglow onset and peak time. Combined with the energy released during the prompt emission, these observations could indicate whether the event satisfies the conditions expected for baryon-loaded relativistic jets.

\begin{figure}[t]
    \centering
    \includegraphics[width=\hsize]{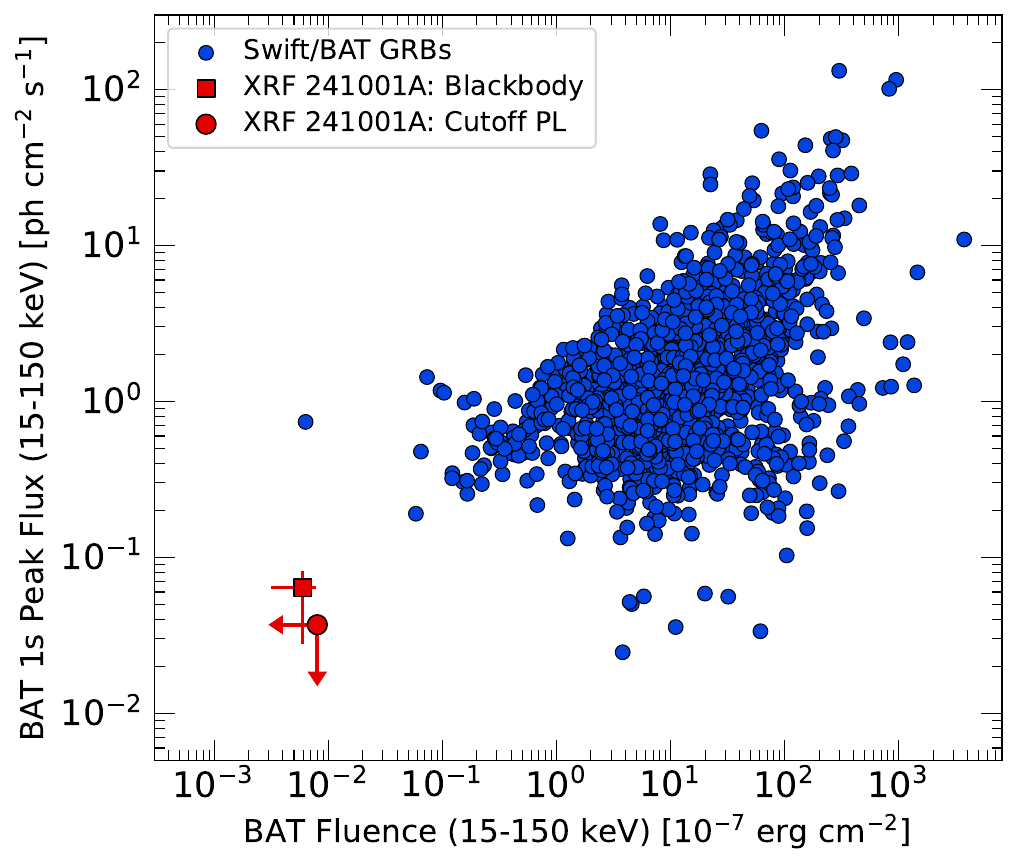}
    \caption{One second peak flux as a function of fluence in the 15–150 keV band for the \Swift/BAT GRB sample (blue) and XRF~241001A (red). Values for the \Swift\ sample are derived assuming a cutoff power-law spectral model. For XRF~241001A, upper limits obtained from the cutoff power-law model are shown as a red circle, and values derived from the blackbody model (best-fit) are shown as a red square for comparison.}
    \label{fig:swiftBATdet}
\end{figure}

By extending sensitivity to the soft energy range, it is expected that \SVOM/ECLAIRs will detect a larger number of X-ray transients and explore the softest part of the GRB population. 
To assess whether XRF~241001A would have been detected by \textit{Swift}, we extrapolate the observed prompt spectrum into the \Swift/BAT 15–150~keV energy range (Table~\ref{tab:xrf241001a_prompt}) and compute the 1~s peak flux and fluence in that band under both blackbody and cutoff power-law assumptions (Table~\ref{tab:xrf241001a_1speak}). These values are compared to those of GRBs in the \Swift/BAT catalog \citep{lien2016a}, for which we adopt measurements derived using a cutoff power-law model. In Fig.~\ref{fig:swiftBATdet}, XRF~241001A clearly occupies a region that is not populated by the \Swift/BAT sample, suggesting that the event would likely have fallen below the typical detection threshold of \Swift/BAT. 
A similar conclusion was proposed for certain eFXTs detected by \EP, which may appear as normal GRBs if observed closer to the jet axis and would have been detected by \Swift/BAT \citep{ren2026a}.
In this context, XRF~241001A suggests that the improved low-energy sensitivity of \SVOM/ECLAIRs might not only enable the detection of off-axis events, as proposed by \citet{ren2026a}, but also detect the intrinsically softer tail of the GRB population, or possibly reveal a distinct population of events, as discussed above for thermal scenarios.

Events similar to XRF~241001A might be a promising avenue for on-axis orphan afterglows detected by ground-based optical surveys such as ZTF \citep{bellm2019a} and, in the near future by the Vera C. Rubin Observatory \citep{ivezic2019a}. While their prompt emission might fall below the detection thresholds of gamma-ray satellites, their afterglows could remain detectable by ZTF up to $z \sim 0.5$ (Fig.~\ref{fig:AG_comparison}) and $z > 1$ by Rubin Observatory. 
Future high-energy constraints from \SVOM/ECLAIRs on these on-axis orphan afterglow candidates will help shed light on the nature of these phenomena.
The systematic multiwavelength follow-up of the soft X-ray transients detected by \SVOM is crucial to fully characterize the population and determine whether these events predominantly arise from successful but intrinsically weak relativistic jets, or whether multiple channels exist. 
A similar event to XRF~241001A observed at a lower redshift (e.g., $z < 0.5$) would provide a prompt emission with a higher S/N to better constrain the nature of the prompt spectrum (thermal versus nonthermal).

The early blue emission observed for XRF~241001A is significantly bluer than is typically observed in GRB afterglows \citep{kann2010a}. Similar features have recently been reported in eFXTs detected by \EP \citep[e.g.,][]{vandalen2025a,eyles-ferris2025a,srinivasaragavan2025b,cotter2026a}. Such emission is difficult to reconcile with standard synchrotron radiation from a relativistic jet and instead suggests the presence of an additional component. Similarly to these \EP events, our modeling of XRF~241001A indicates the presence of an additional thermal emission. Its origin remains challenging to investigate with our dataset, providing poor constraints on the spectral and temporal evolution of this component. Proposed scenarios include a shocked cocoon \citep{izzo2019a}, either produced by a jet choked in its stellar envelope \citep{nakar2017a} or by the interaction of the jet with an extended circumstellar medium \citep{nakar2015a}.
The similarities between the early emission properties of XRF~241001A and eFXTs detected by \EP suggest that early blue components may be a common feature of soft X-ray transients and might indicate a link regarding their production mechanisms. 
XRF~241001A exhibited remarkably faint X-ray and optical afterglow emissions compared to the bulk of the long GRB population. At $T_0 + 11$~hr, it lies within the faintest $\sim$5\% of events (Fig.~\ref{fig:AG_comparison}). Under these circumstances, the detection of an additional emission might indicate that similar features could be present regularly in other events but are typically outshone by brighter afterglows and therefore remain largely undetected. 

XRF~241001A occurred during the \SVOM commissioning phase, at a time when the automatic slew capability was not yet enabled, resulting in a significant delay in the first \SVOM/VT observations. 
With SVOM now fully operational, these events benefit from early optical coverage in two simultaneous bands. This provides prompt color information on the afterglow emission and will offer a way to assess whether the early blue excess is a common feature of soft transients and to investigate its origin.

\section{Conclusions}
\label{sec:Conclusions}
We have presented the detection and multiwavelength analysis of XRF~241001A detected by \SVOM/ECLAIRs. Its prompt emission is dominated by soft X-ray photons below $\sim$20~keV, and it can be modeled by either thermal or nonthermal models with $E_{\rm peak} < 10$~keV. The limited S/N prevents a clear discrimination between models, but our analyses point toward a nonthermal interpretation.
At $z = 0.5728$, it has an isotropic energy of $E_{\rm iso} \sim 8\times10^{49}$~erg, placing it among the least energetic GRBs detected at similar redshifts and in a sparsely populated region of the Amati plane. The prompt emission of XRF~241001A lies well outside the distribution of long GRBs detected by \Swift, occupying a region of low peak flux and low fluence, and it would likely have remained undetected by \Swift/BAT.

The afterglow emission of XRF~241001A is notably faint in both X-rays and optical bands, placing it among the faintest $\sim$5\% of the long GRB distribution.
The afterglow modeling indicates that the emission can be well described by a relativistic jet viewed on-axis interacting with a moderately dense circumburst medium. Together with the prompt emission, this implies a low gamma-ray efficiency.
The VLT/X-shooter observations show the presence of an early blue optical emission, similar to that observed in some eFXTs detected by \EP at a similar epoch. Its origin remains uncertain, but its thermal nature is difficult to reconcile with the synchrotron emission produced by a relativistic jet.
At later times, \JWST/NIRSpec and \SVOM/VT observations reveal the emergence of an SN Ic-BL with properties similar to GRB/SNe, firmly establishing a massive star progenitor.
The host galaxy of XRF~241001A is a faint sub-$L^*$ low-metallicity galaxy with ongoing star formation, consistent with the long GRB host galaxies observed at similar redshifts.

The prompt afterglow and SN properties of XRF~241001A support the view that at least a fraction of the XRF population belongs to the soft low-luminosity tail of the long GRB distribution. The event XRF~241001A remarkably demonstrates the enhanced sensitivity of \SVOM/ECLAIRs to softer and fainter high-energy transients, revealing a population that has been largely underrepresented in previous GRB samples.

\section*{Data availability}
The JWST/NIRSpec data used in this work are publicly available through WISeREP at \url{https://www.wiserep.org/object/31052} and can be found in MAST at \url{https://doi.org/10.17909/nwnm-y263}.

\begin{acknowledgements}
    We thank the anonymous referee for the careful reading of this manuscript and for constructive comments that helped improve the quality and clarity of this work.
    BS, VB, SDV and ELF acknowledge the support of the French Agence Nationale de la Recherche (ANR), under grant ANR-23-CE31-0011 (project PEGaSUS).
    BPG acknowledges support from STFC grant No. ST/Y002253/1 and from the Leverhulme Trust grant No. RPG-2024-117.
    DBM is funded by the European Union (ERC, HEAVYMETAL, 101071865). Views and opinions expressed are, however, those of the authors only and do not necessarily reflect those of the European Union or the European Research Council. Neither the European Union nor the granting authority can be held responsible for them. The Cosmic Dawn Center (DAWN) is funded by the Danish National Research Foundation under grant DNRF140. 
    AS acknowledges financial support from the Centre national d’études spatiales (CNES), France (ROR: \url{https://ror.org/04h1h0y33}) within the framework of the SVOM mission.
    The authors are thankful for support from the National Key R\&D Program of China (grant Nos. 2024YFA161170* and 2024YFA1611700).This work is supported by the Strategic Priority Research Program of the Chinese Academy of Sciences, Grant No.XDB055040, XDB0550100.
    NRT is supported by STFC grants ST/W000857/1 and UKRI1200. 
    YFH was supported by the National Key R\&D Program of China (2021YFA0718500) and by the Xinjiang Tianchi Program.
    The authors thanks P. Hello for the operations of GRANDMA as well as the GRANDMA core team, the reviewers N. Guessoum and N. Kochiashvili and M. Freeberg for the observations under Kilonova-catcher with the iT72 télescope. SA and AK thanks all the TAROT operational team as C. Limonta, C. Durroux and Q. André.
    The Space-based multiband astronomical Variable Objects Monitor (SVOM) is a joint Chinese-French mission led by the Chinese National Space Administration (CNSA), the French Space Agency (CNES), and the Chinese Academy of Sciences (CAS). We gratefully acknowledge the unwavering support of NSSC, IAMCAS, XIOPM, NAOC, IHEP, CNES, CEA, and CNRS. This work is supported by the National Natural Science Foundation of China (grant Nos. 12494571 and 12494570, 12494573, 12133003). 
    This work is based in part on observations made with the NASA/ESA/CSA James Webb Space Telescope. The data were obtained from the Mikulski Archive for Space Telescopes at the Space Telescope Science Institute, which is operated by the Association of Universities for Research in Astronomy, Inc., under NASA contract NAS 5-03127 for JWST. These observations are associated with program \#6133.
    Partly based on observations carried out under ESO prog. ID 114.27PZ (PIs: N. R. Tanvir, S. D. Vergani. D. B. Malesani) with the X-shooter spectrograph installed at the Cassegrain focus of the Very Large Telescope (VLT), Unit 3 (Melipal), operated by the European Southern Observatory (ESO) on Cerro Paranal, Chile.
    This paper includes data gathered with the 6.5 meter Magellan Telescopes located at Las Campanas Observatory, Chile. 
    The Australia Telescope Compact Array is part of the Australia Telescope National Facility (\url{https://ror.org/05qajvd42}) which is funded by the Australian Government for operation as a National Facility managed by CSIRO.
    We acknowledge the Gomeroi people as the Traditional Owners of the Observatory site.
    This work made use of data supplied by the UK Swift Science Data Centre at the University of Leicester.
\end{acknowledgements}

\bibliographystyle{aa}
\bibliography{ref}

\appendix
\section{Additional materials}
\label{app:additional_materials}
In this appendix, we present additional details on the prompt and afterglow analysis of XRF~241001A. 
The results of the time-integrated prompt spectral fits obtained with different spectral models 
(blackbody, power law, broken power law, and cutoff power law) are summarized in 
Table~\ref{tab:xrf241001a_prompt} and the one-second peak spectral analysis in the 15--150\,keV band is reported in Table~\ref{tab:xrf241001a_1speak}. 
The prompt spectral analysis was performed using the following user-defined energy bands: 
Band~1 (3.9--5.1\,keV), 
Band~2 (5.1--6.1\,keV), 
Band~3 (6.1--8.1\,keV), 
Band~4 (8.1--10.1\,keV), 
Band~5 (10.1--13.1\,keV), 
Band~6 (13.1--17.1\,keV), 
Band~7 (17.1--22.1\,keV), 
and Band~8 (22.1--28.1\,keV).

The results of the X-ray afterglow spectral analysis are listed in Table~\ref{tab:x-ray_spec}. 
The complete optical photometry used in this work is reported in Table~\ref{table:photometry}, and the radio observations are given in Table~\ref{table:radio}. 
The priors adopted and the resulting posterior constraints for the afterglow (top-hat jet) and supernova (Arnett) models are summarized in Table~\ref{tab:redback_params}. 
The emission-line flux measurements of the host galaxy are presented in Table~\ref{tab:host_em_line_flux}. 

\begin{figure*}[t]
\centering
    \includegraphics[width=13.5cm]{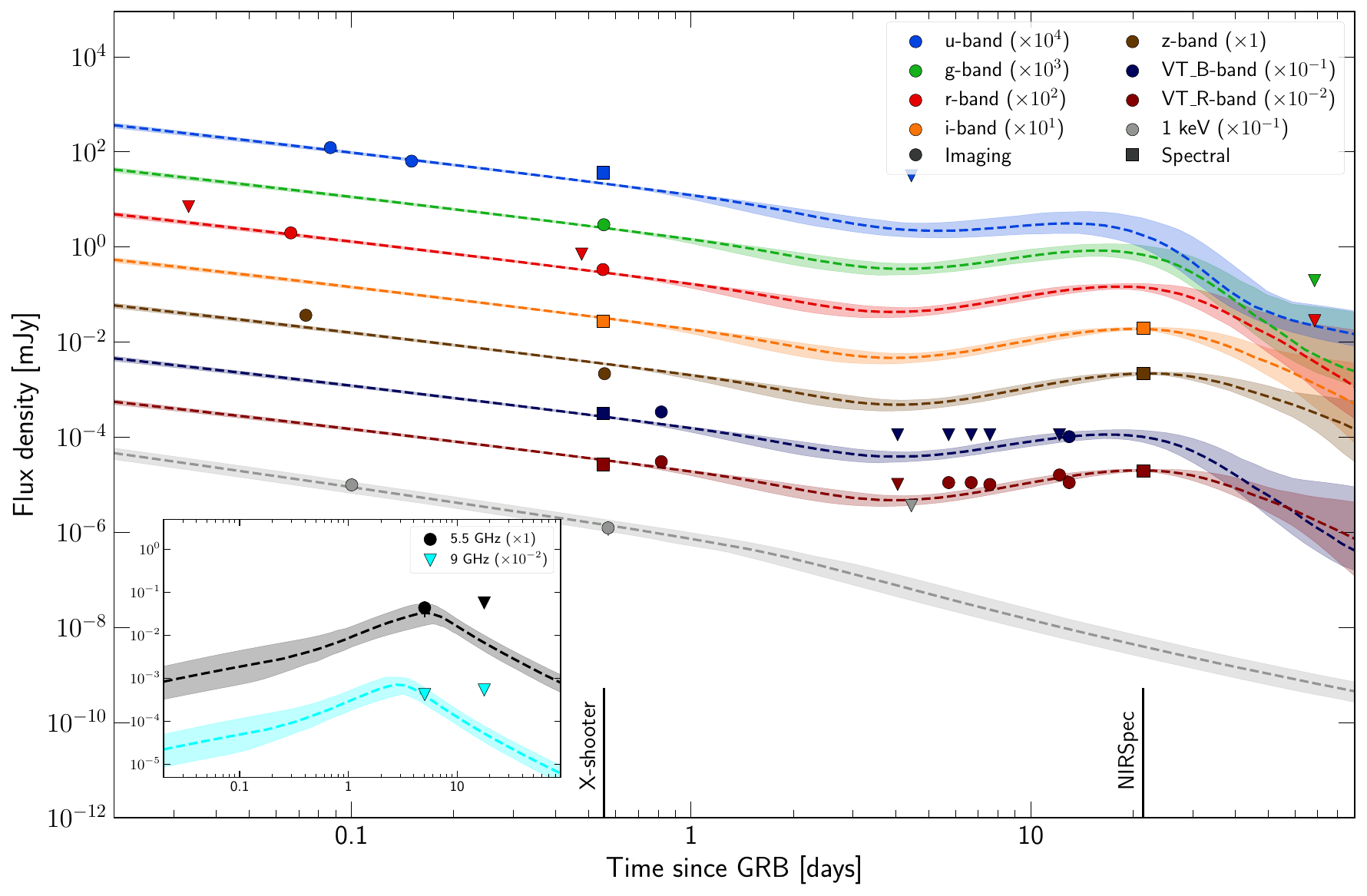}
    \caption{Same as Fig.~\ref{fig:redback_fit_model2} but showing the best-fit model (top-hat jet + Arnett) using the full dataset, including the thermal component.}
    \label{fig:redback_fit_model1}
\end{figure*}

\begin{figure*}[t]
\centering
    \includegraphics[width=13cm]{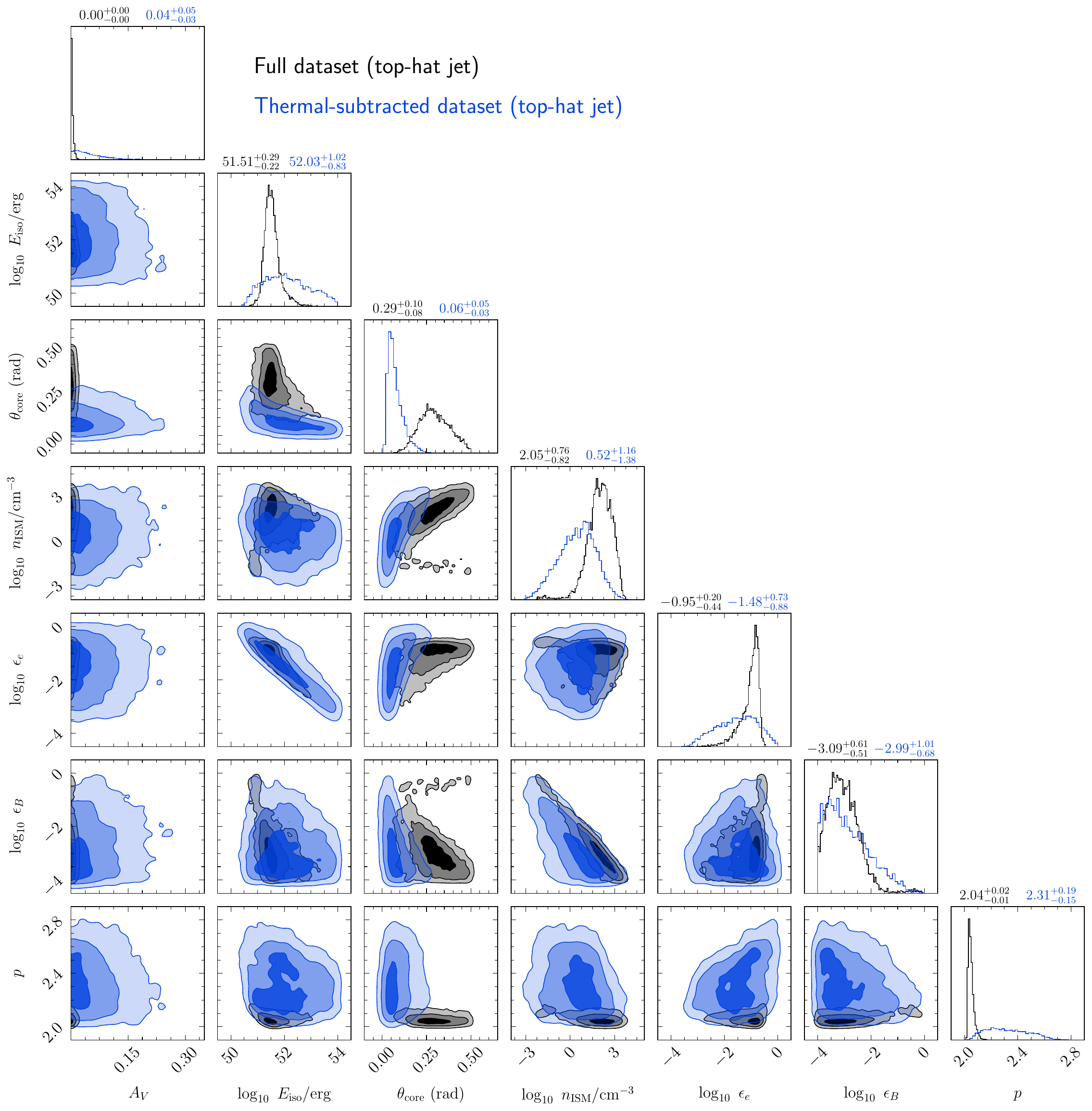}
      \caption{Corner plot of the posterior distributions for the afterglow parameters inferred from the top-hat jet modeling of observations obtained within the first day after the trigger. The posterior distributions derived from the dataset including the thermal component are shown in black, while those obtained after subtracting the thermal component are shown in blue.}
      \label{fig:redback_corner_AG}
\end{figure*}

\begin{figure*}[t]
\centering
    \includegraphics[width=13cm]{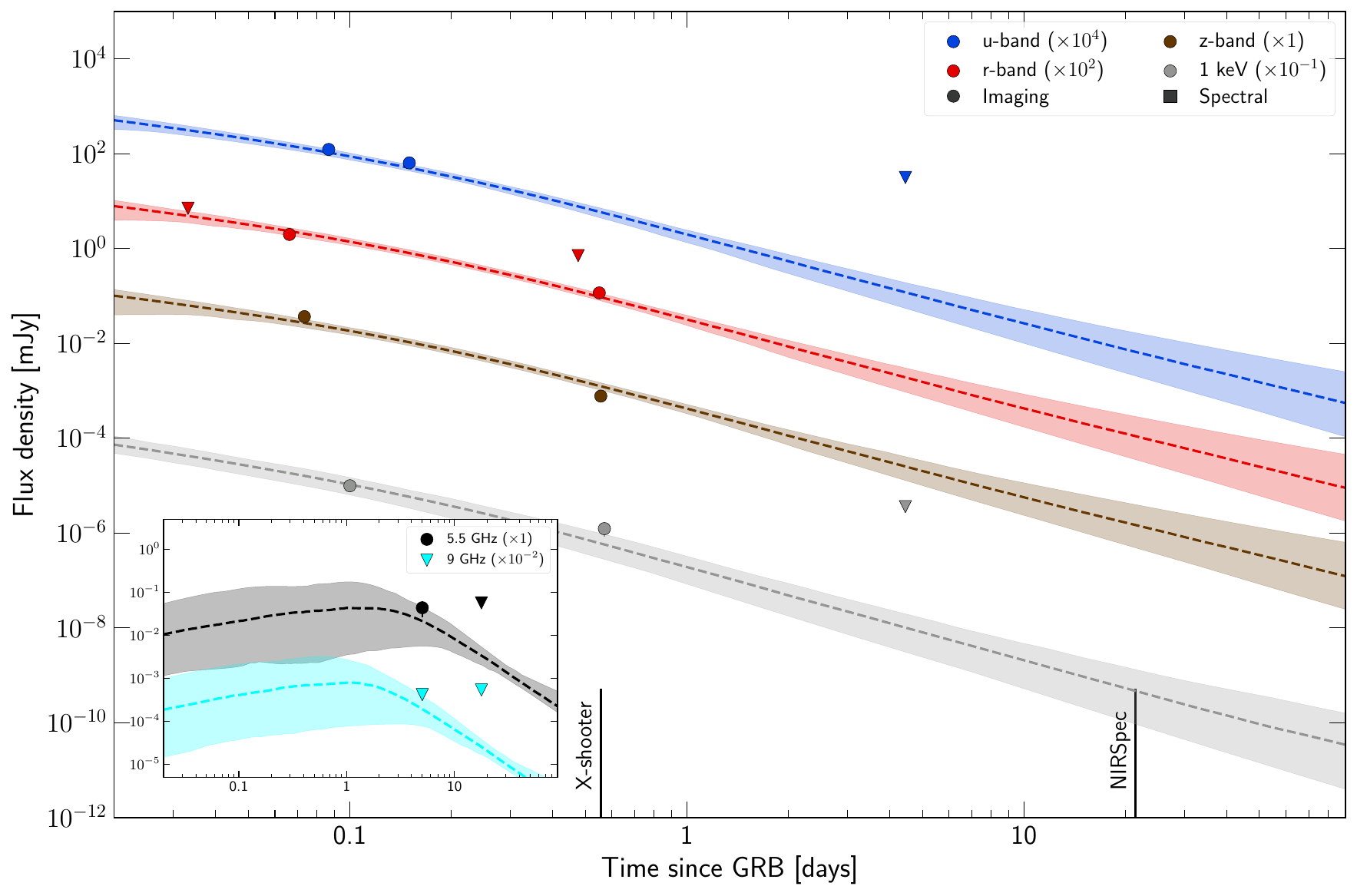}
    \caption{Similar to Fig.~\ref{fig:redback_fit_model2} but showing the best-fit Gaussian jet model determined for the observations acquired within the first day after the trigger and subtracted from the thermal component (scenario 2).}
    \label{fig:redback_fit_model_gauss}
\end{figure*}

\begin{figure*}[t]
\centering
    \includegraphics[width=13cm]{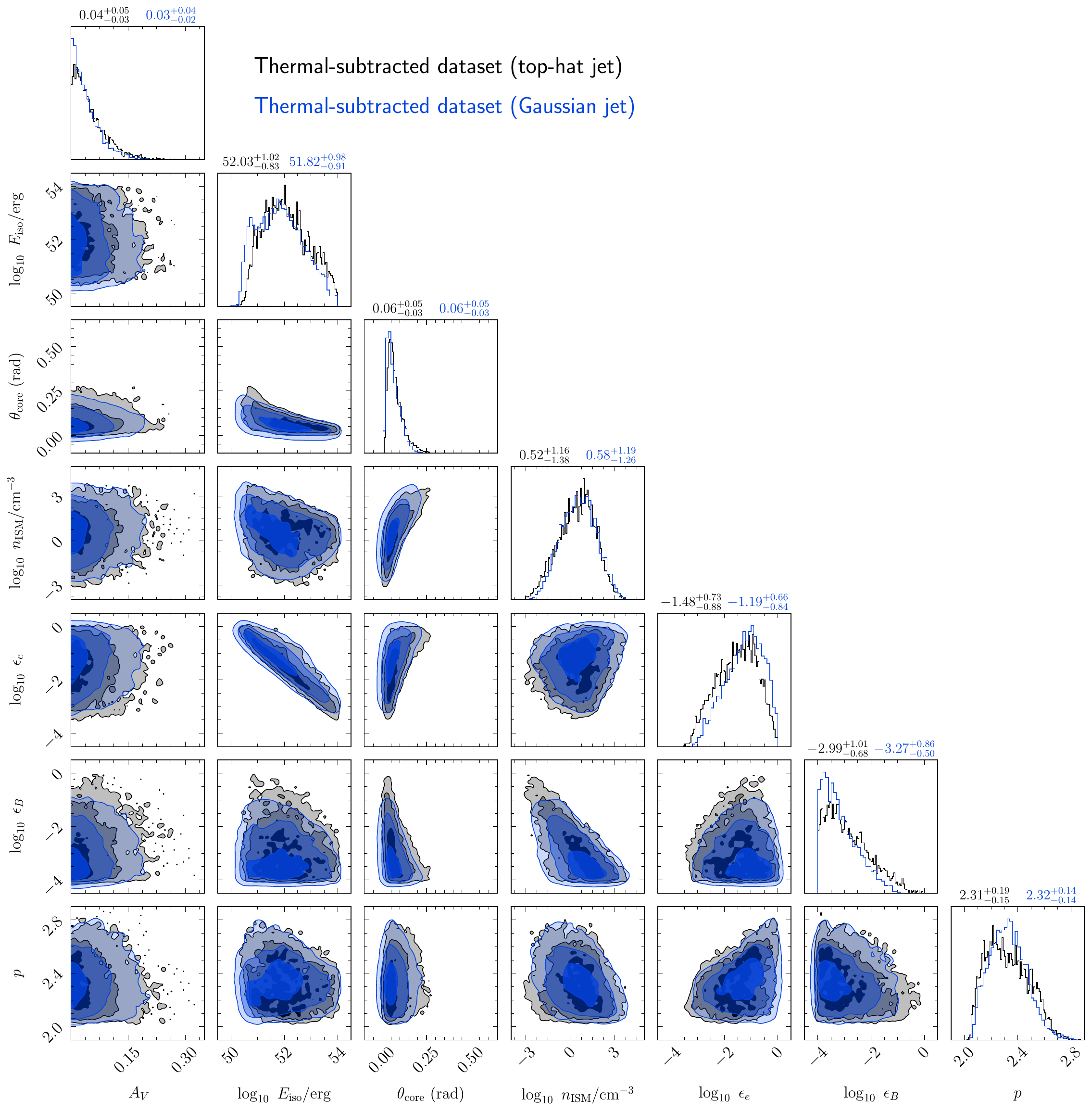}
    \caption{Same as Fig.~\ref{fig:redback_corner_AG} but showing the posterior distributions returned by the top-hat (black) and Gaussian (blue) jet models for the observations acquired within the first day after the trigger and subtracted from the thermal component (scenario 2).}
    \label{fig:redback_corner_AG_tophat_gaussian}
\end{figure*}

\begin{figure*}[t]
\centering
    \includegraphics[width=11cm]{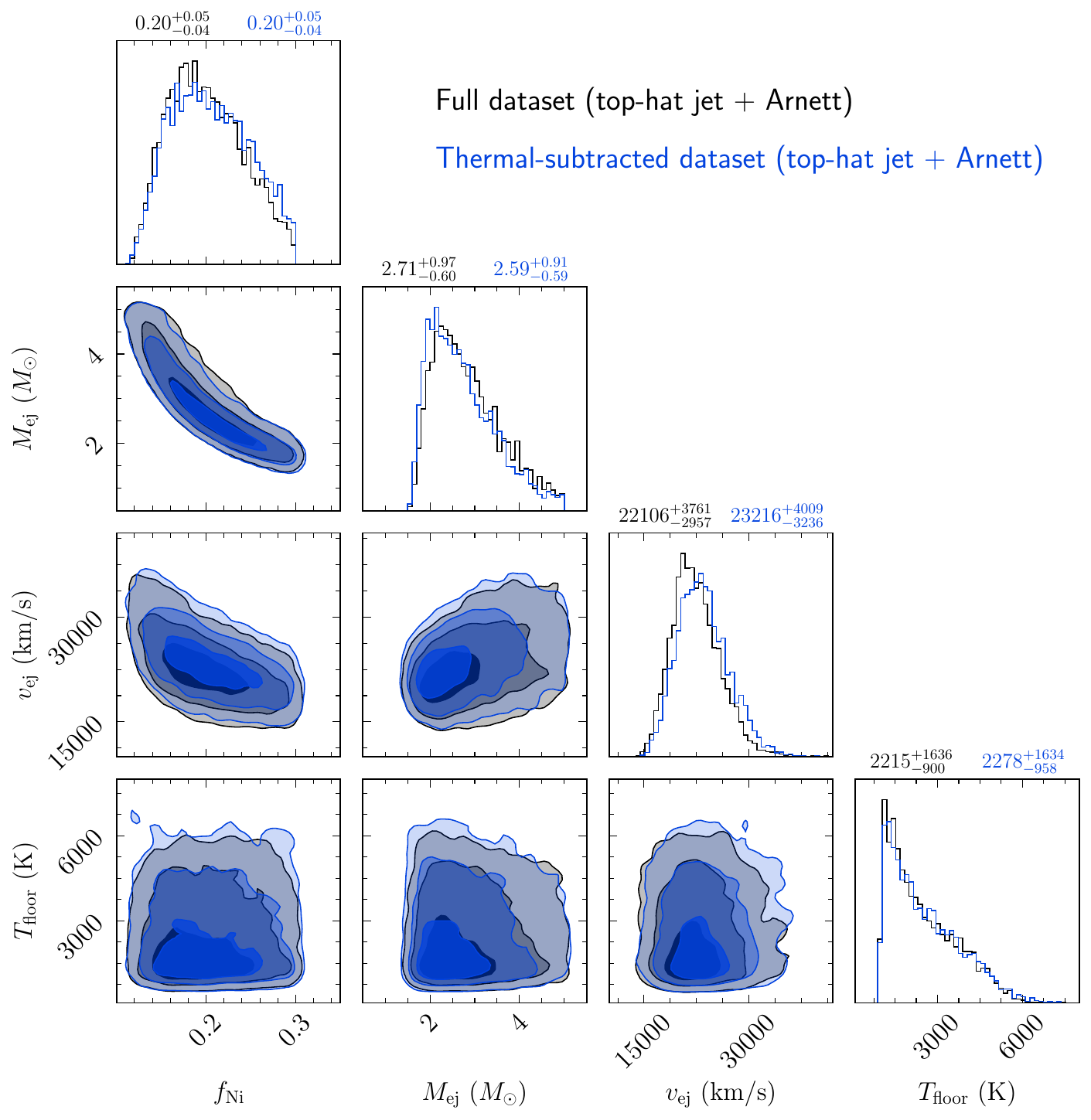}
    \caption{Same as Fig.~\ref{fig:redback_corner_AG} but for the SN parameters obtained from the joint afterglow (top-hat jet) and SN (Arnett) modeling using the full dataset (black) and the dataset after subtracting the thermal component (blue).}
    \label{fig:redback_corner_SN}
\end{figure*}

The best-fit light curve model obtained using the full dataset is shown in 
Fig.~\ref{fig:redback_fit_model1} 
and the Gaussian jet fit obtained with the thermal-subtracted dataset is visible in Fig.~\ref{fig:redback_fit_model_gauss}.
The posterior distributions are presented in Fig.~\ref{fig:redback_corner_AG} for the top-hat afterglow model for both datasets, in Fig.~\ref{fig:redback_corner_AG_tophat_gaussian} for the top-hat versus Gaussian jet models using the thermal-subtracted dataset, and in Fig.~\ref{fig:redback_corner_SN} for the supernova parameters considering both datasets.

\renewcommand{\arraystretch}{1.35}
\begin{table}[h]
\centering
\caption{Results of the time-integrated prompt spectral analysis.}
\label{tab:xrf241001a_prompt}
\begin{tabular}{ll}
\hline

\multicolumn{2}{c}{Cutoff Power-Law model} \\
\hline
$\chi^2/\mathrm{d.o.f.}$ & 3.43/4 \\
$\Gamma$ & $-0.53^{+1.65}_{-2.25}$ \\
Norm & $2.13^{+2.92}_{-1.60}$ \\
$E_{\rm cut}$ & $2.48^{+2.57}_{-1.00}$ keV \\
$F_{4-20\,{\rm keV}}$ &
$(1.96^{+0.07}_{-0.08})\times10^{-8}$ erg\,cm$^{-2}$\,s$^{-1}$ \\
$F_{4-120\,{\rm keV}}$ &
$(1.99^{+0.10}_{-0.10})\times10^{-8}$ erg\,cm$^{-2}$\,s$^{-1}$ \\
$F_{15-150\,{\rm keV}}$ &
$(1.25^{+0.82}_{-0.61})\times10^{-9}$ erg\,cm$^{-2}$\,s$^{-1}$ \\
$E_{\rm peak}$ & $6.27^{+0.87}_{-1.07}$ keV \\
$L_{\rm iso}$ & $(4.48^{+1.05}_{-0.88})\times10^{49}$ erg\,s$^{-1}$ \\
$E_{\rm iso}$ & $(8.95^{+2.10}_{-1.76})\times10^{49}$ erg \\
\hline

\multicolumn{2}{c}{Broken Power-Law model} \\
\hline
$\chi^2/\mathrm{d.o.f.}$ & 2.84/3 \\
$\Gamma_1$ & $1.90^{+0.46}_{-0.56}$ \\
$E_{\rm break}$ & $9.34^{+1.19}_{-1.36}$ keV \\
$\Gamma_2$ & $5.13^{+3.71}_{-1.46}$ \\
Norm & $8.6^{+11.2}_{-5.4}$ \\
$F_{4-20\,{\rm keV}}$ &
$(1.94^{+0.10}_{-1.48})\times10^{-8}$ erg\,cm$^{-2}$\,s$^{-1}$ \\
$F_{4-120\,{\rm keV}}$ &
$(2.00^{+0.32}_{-1.72})\times10^{-8}$ erg\,cm$^{-2}$\,s$^{-1}$ \\
$F_{15-150\,{\rm keV}}$ &
$(1.36^{+3.72}_{-1.23})\times10^{-9}$ erg\,cm$^{-2}$\,s$^{-1}$ \\
\hline
\end{tabular}
\end{table}

\begin{table}
\centering
\begin{tabular}{ll}
\hline

\multicolumn{2}{c}{Power-Law model} \\
\hline
$\chi^2/\mathrm{d.o.f.}$ & 8.45/5 \\
$\Gamma$ & $2.73^{+0.24}_{-0.23}$ \\
Norm & $37.6^{+21.9}_{-13.7}$ \\
$F_{4-20\,{\rm keV}}$ &
$(2.06^{+0.08}_{-0.49})\times10^{-8}$ erg\,cm$^{-2}$\,s$^{-1}$ \\
$F_{4-120\,{\rm keV}}$ &
$(2.72^{+0.21}_{-0.78})\times10^{-8}$ erg\,cm$^{-2}$\,s$^{-1}$ \\
$F_{15-150\,{\rm keV}}$ &
$(9.19^{+2.55}_{-4.03})\times10^{-9}$ erg\,cm$^{-2}$\,s$^{-1}$ \\
\hline

\multicolumn{2}{c}{Blackbody model} \\
\hline
$\chi^2/\mathrm{d.o.f.}$ & 3.72/5 \\
$kT$ & $1.72^{+0.16}_{-0.15}$ keV \\
Norm & $269.7^{+129.1}_{-86.0}$ \\
Radius & $2.20^{+0.40}_{-0.48}\times10^{11}$ cm \\
$F_{4-20\,{\rm keV}}$ &
$(1.89^{+0.11}_{-0.40})\times10^{-8}$ erg\,cm$^{-2}$\,s$^{-1}$ \\
$F_{4-120\,{\rm keV}}$ &
$(1.90^{+0.10}_{-0.40})\times10^{-8}$ erg\,cm$^{-2}$\,s$^{-1}$ \\
$F_{15-150\,{\rm keV}}$ &
$(5.95^{+1.92}_{-2.82})\times10^{-10}$ erg\,cm$^{-2}$\,s$^{-1}$ \\
$E_{\rm peak}$ & $6.73^{+0.61}_{-0.57}$ keV \\
$L_{\rm iso}$ & $(3.58^{+0.20}_{-0.68})\times10^{49}$ erg\,s$^{-1}$ \\
$E_{\rm iso}$ & $(7.15^{+0.83}_{-1.69})\times10^{49}$ erg \\
\hline
\end{tabular}
\tablefoot{
The $E_{\rm peak}$, $L_{\rm iso}$, and $E_{\rm iso}$ for the cutoff power-law and blackbody models are computed from the best-fit model.}
\end{table}

\renewcommand{\arraystretch}{1.35}
\begin{table}
\centering
\caption{Results of the 1~s peak spectral analysis in the 15--150~keV band.}
\label{tab:xrf241001a_1speak}
\begin{tabular}{ll}
\hline

\multicolumn{2}{c}{Cutoff Power-Law model} \\
\hline
$\chi^2/\mathrm{d.o.f.}$ & 4.57/4 \\
$\Gamma$ & $2.0 \pm 2.4$ \\
$E_{\rm cut}$ & $1.9 \pm 1.2$ keV \\
Norm & $0.2 \pm 0.6$ \\
${\rm Flux}_{15-150\,{\rm keV}}$ &
<$0.04$ ph\,cm$^{-2}$\,s$^{-1}$ \\
\hline

\multicolumn{2}{c}{Broken Power-Law model} \\
\hline
$\chi^2/\mathrm{d.o.f.}$ & 2.75/3 \\
$\Gamma_1$ & $1.01^{+0.55}_{-0.62}$ \\
$E_{\rm break}$ & $10.3 \pm 1.1$ keV \\
$\Gamma_2$ & $8.9 \pm 9.4$ \\
Norm & $<$$6.0$  \\
${\rm Flux}_{15-150\,{\rm keV}}$ &
<$1.6$ ph\,cm$^{-2}$\,s$^{-1}$ \\
\hline

\multicolumn{2}{c}{Power-Law model} \\
\hline
$\chi^2/\mathrm{d.o.f.}$ & 14.0/7 \\
$\Gamma$ & $2.4 \pm 0.3$ \\
Norm & $23.9 \pm 15.6$ \\
${\rm Flux}_{15-150\,{\rm keV}}$ &
<$0.56$ ph\,cm$^{-2}$\,s$^{-1}$ \\
\hline

\multicolumn{2}{c}{Blackbody model} \\
\hline
$\chi^2/\mathrm{d.o.f.}$ & 4.76/5 \\
$kT$ & $1.99 \pm 0.20$ keV \\
Norm & $200.00 \pm 81.25$ \\
${\rm Flux}_{15-150\,{\rm keV}}$ &
$0.06 \pm 0.03 $ ph\,cm$^{-2}$\,s$^{-1}$ \\
\hline

\end{tabular}
\end{table}

\renewcommand{\arraystretch}{1.2}
\begin{table*}
    \caption{Results of the X-ray afterglow spectral analysis.}
    \centering
    \begin{tabular}{ccccccc}
         \hline
         Instrument & $T-T_{0}^{(1)}$ & $N^{\rm Gal}_{H,X}{}^{(2)}$ & $N^{\rm host}_{H,X}$ & $\Gamma$ & Flux & Unabsorbed Flux\\ \cline{6-7}
                    & (seconds) & ($10^{20}\,\mathrm{cm}^{-2}$) & ($10^{22}\,\mathrm{cm}^{-2}$) &   & \multicolumn{2}{c}{($10^{-13}\,\mathrm{erg\,cm^{-2}\,s^{-1}}$)} \\
         \hline
         \Swift/XRT & 6536 -- 13659 & 1.28 & $0.57^{+0.54}_{-0.36}$ & $1.70^{+0.45}_{-0.37}$ & $7.64^{+1.78}_{-2.52}$ & $9.54^{+2.47}_{-1.96}$  \\
          \EP/FXT & 44879 -- 53628 & 1.28 & $0.61^{+0.32}_{-0.28}$ & $2.30^{+0.65}_{-0.60}$ & $0.81^{+0.14}_{-0.25}$ & $1.20^{+0.48}_{-0.31}$  \\
          \Swift/XRT & 370635 -- 456166 & 1.28 & $0.57^{(\dagger)}$ & $1.70^{(\dagger)}$ & <3.09 & <3.50 \\
         \hline
          \multirow{2}{*}{\Swift/XRT + \EP/FXT} & \multirow{2}{*}{6536 -- 53628} & \multirow{2}{*}{1.28} & \multirow{2}{*}{$0.52^{+0.41}_{-0.30}$} & \multirow{2}{*}{$1.80^{+0.37}_{-0.30}$} &  $6.96^{+1.19}_{-1.43}$ (XRT)  & $8.71^{+1.76}_{-1.63}$ (XRT) \\
            &  &   &   &  &  $0.99^{+0.22}_{-0.31}$ (FXT)  & $1.20^{+0.30}_{-0.26}$ (FXT) \\
          \hline
    \end{tabular}
    \tablefoot{Fluxes are computed in the 0.3--10~keV energy range and upper limits are given at the 3$\sigma$ confidence level. \\
               $^{(1)}$$T_{0} = \mathrm{2024\text{-}10\text{-}01T17{:}08{:}57.74}$ UT.\\
               $^{(2)}$ Galactic neutral atomic hydrogen (\ion{H}{i}) column density, from $N_{\ion{H}{i}}$ map of \cite{bekhti2016a}.\\
               $^{(\dagger)}$ Values fixed during the fit. 
               }
    \label{tab:x-ray_spec}
\end{table*}

\begin{table*}
\caption{Optical observations of XRF~241001A/SN~2024aii.}
\label{table:photometry}
\centering
\begin{tabular}{ccccc} 
\hline
$T-T_{0}$ (s) & Telescope/Instrument & Band & Exposure time (s) & AB magnitude \\ 
\hline
2863 & TRE & $r$ & 19$\times$120 & >19.3 \\
5725 & LCO & $r$ & 3$\times$120 & $20.68 \pm 0.19$ \\
6339 & LCO & $z$ & 5$\times$120 & $20.00 \pm 0.23$ \\
7486 & \Swift/UVOT & $u$ & 1875 & $21.22 \pm 0.15^{(1)}$ \\
12983 & \Swift/UVOT & $u$ & 1324 & $21.93 \pm 0.24^{(1)}$ \\
41145 & KCN/iT72 & $r$ & 18$\times$180 & >21.8 \\
47793 & VLT/X-shooter & $g$ & 3$\times$60 & $22.76 \pm 0.10$ \\
47526 & VLT/X-shooter & $r$ & 3$\times$60 & $22.61 \pm 0.08$ \\
48026 & VLT/X-shooter & $z$ & 3$\times$60 & $23.07 \pm 0.22$ \\
47526 & VLT/X-shooter & $u$ & -- & $22.48 \pm 0.05^{(2)}$ \\
47526 & VLT/X-shooter & $i$ & -- & $22.81 \pm 0.06^{(2)}$ \\
47526 & VLT/X-shooter & $VT\_B$ & -- & $22.65 \pm 0.08^{(2)}$ \\
47526 & VLT/X-shooter & $VT\_R$ & -- & $22.82 \pm 0.06^{(2)}$ \\
70578 & SVOM/VT & $VT\_B$ & 195$\times$20 & $22.6 \pm 0.2$ \\
70578 & SVOM/VT & $VT\_R$ & 194$\times$20 & $22.7 \pm 0.2$ \\
384710 & \Swift/UVOT & $u$ & 4126 & >22.7 \\
350354 & SVOM/VT & $VT\_B$ & 158$\times$20 & >23.8 \\
350354 & SVOM/VT & $VT\_R$ & 189$\times$20 & >23.9 \\
495169 & SVOM/VT & $VT\_B$ & 225$\times$20 & >23.8 \\
495169 & SVOM/VT & $VT\_R$ & 248$\times$20 & $23.8 \pm 0.3$ \\
576914 & SVOM/VT & $VT\_B$ & 176$\times$20 & >23.8 \\
576914 & SVOM/VT & $VT\_R$ & 253$\times$20 & $23.8 \pm 0.3$ \\
654554 & SVOM/VT & $VT\_B$ & 176$\times$20 & >23.8 \\
654554 & SVOM/VT & $VT\_R$ & 252$\times$20 & $23.9 \pm 0.3$ \\
1049402 & SVOM/VT & $VT\_B$ & 282$\times$20 & >23.8 \\
1049402 & SVOM/VT & $VT\_R$ & 330$\times$20 & $23.4 \pm 0.25$ \\
1121036 & SVOM/VT & $VT\_B$ & 668$\times$20 & $23.9 \pm 0.3$ \\
1121036 & SVOM/VT & $VT\_R$ & 668$\times$20 & $23.8 \pm 0.3$ \\
1852664 & JWST/NIRSpec & $i$ & -- & $23.16 \pm 0.06^{(2)}$ \\
1852664 & JWST/NIRSpec & $z$ & -- & $23.05 \pm 0.05^{(2)}$ \\
1852664 & JWST/NIRSpec & $VT\_R$ & -- & $23.16 \pm 0.05^{(2)}$ \\
5918552 & Magellan/LDSS & $g$ & 4$\times$300 & >25.7 \\
5920114 & Magellan/LDSS & $r$ & 5$\times$300 & >25.3 \\
\hline
\end{tabular}
\tablefoot{Magnitudes are given in AB system and are not corrected for Galactic extinction. Rows without exposure times correspond to synthetic magnitudes derived from spectroscopy data. \\
$^{(1)}$\cite{breeveld2024a}. \\
$^{(2)}$Synthetic magnitude derived from spectroscopy data.}
\end{table*}

\begin{table}
\caption{Radio observations of XRF~241001A.}      
\label{table:radio}
\centering
\begin{tabular}{cccc} 
\hline
$T-T_{0}$ (s) & Telescope & Frequency (GHz) & Flux (mJy) \\ 
\hline
434745 & ATCA & 5.5 & $0.044 \pm 0.018$ \\
434745 & ATCA & 9.0 & $<$0.042 \\
1536560 & ATCA & 5.5 & $<$0.057 \\
1536560 & ATCA & 9.0 & $<$0.054 \\
\hline
\end{tabular}
\tablefoot{Upper limits correspond to 3$\sigma$ non-detections.}
\end{table}

\renewcommand{\arraystretch}{1.3}
\begin{table}
\vspace*{3.5cm}
\centering
\makebox[\textwidth][c]{%
\parbox{\textwidth}{%
\caption{Priors and posterior values from the afterglow and supernova modeling of XRF~241001A/SN~2024aiiq.}
\label{tab:redback_params}
}%
}
\makebox[\textwidth][c]{%
\begin{tabular}{lcccccc}
\hline
Parameter & Prior & \multicolumn{4}{c}{Posterior} \\
\cline{3-4}\cline{5-6}
 &  & \multicolumn{2}{c}{Full dataset} 
    & \multicolumn{2}{c}{Thermal-subtracted dataset} \\
\cline{3-6}
&  & Top-hat & Top-hat + Arnett & Top-hat & Top-hat + Arnett \\
\hline

$z$ & 0.57 & 0.57 & 0.57 & 0.57 & 0.57 \\
$\theta_v$ (rad) & 0 & 0 & 0 & 0 & 0 \\
$\Gamma_0$ & 100 & 100 & 100 & 100 & 100 \\
$\kappa$ & 0.07 & & 0.07 & & 0.07 \\
$\kappa_\gamma$ & 0.3 & & 0.3 & & 0.3 \\
$k_{\rm sin}$ & 1 & & 1 & & 1 \\
$A_V^{\rm GRB}$ (mag) & 0 & & 0 & & 0  \\
\hline

$\log E_0$ (erg)
& $\mathcal{U}(44,54)$
& $51.51^{+0.29}_{-0.22}$
& $51.40^{+0.15}_{-0.10}$
& $52.00^{+1.00}_{-0.83}$
& $52.00^{+0.52}_{-0.50}$ \\

$\log \epsilon_e$
& $\mathcal{U}(-5,0)$
& $-0.95^{+0.20}_{-0.44}$
& $-1.14^{+0.17}_{-0.14}$
& $-1.48^{+0.73}_{-0.88}$
& $-1.58^{+0.41}_{-0.44}$ \\

$\log \epsilon_B$
& $\mathcal{U}(-5,0)$
& $-3.09^{+0.61}_{-0.51}$
& $-3.36^{+0.27}_{-0.16}$
& $-2.99^{+1.01}_{-0.68}$
& $-3.07^{+0.52}_{-0.39}$ \\

$p$
& $\mathcal{U}(2,3)$
& $2.04^{+0.02}_{-0.01}$
& $2.04^{+0.02}_{-0.01}$
& $2.31^{+0.19}_{-0.15}$
& $2.33^{+0.10}_{-0.12}$ \\

$\log n_0$ (cm$^{-3}$)
& $\mathcal{U}(-5,2)$
& $2.05^{+0.76}_{-0.82}$
& $2.05^{+0.30}_{-0.39}$
& $0.52^{+1.16}_{-1.38}$
& $0.21^{+0.73}_{-0.63}$ \\

$\theta_c$ (rad)
& $\mathcal{U}(0.01,0.5)$
& $0.29^{+0.10}_{-0.08}$
& $0.28^{+0.05}_{-0.04}$
& $0.06^{+0.05}_{-0.03}$
& $0.06^{+0.02}_{-0.02}$ \\

$f_{\rm Ni}$
& $\log\mathcal{U}(0.1,0.6)$
& 
& $0.20^{+0.05}_{-0.04}$
& 
& $0.20^{+0.05}_{-0.04}$ \\

$M_{\rm ej}$ ($M_\odot$)
& $\log\mathcal{U}(1,8)$
& 
& $2.71^{+0.97}_{-0.60}$
& 
& $2.59^{+0.91}_{-0.59}$ \\

$v_{\rm ej}$ (km s$^{-1}$)
& $\log\mathcal{U}(10^3,5\times10^4)$
& 
& $(2.21^{+0.38}_{-0.30})\times10^4$
& 
& $(2.32^{+0.40}_{-0.32})\times10^4$ \\

$T_{\rm floor}$ (K)
& $\log\mathcal{U}(10^3,10^5)$
& 
& $(2.22^{+1.64}_{-0.90})\times10^3$
& 
& $(2.28^{+1.63}_{-0.96})\times10^3$ \\

$A_V^{\rm GRB}$ (mag)
& $\mathcal{U}(0,1)$
& $0.00^{+0.00}_{-0.00}$
&
& $0.04^{+0.05}_{-0.03}$
&  \\
\hline
$\log Z$
& --
& 2.50
& 71.37
& 38.09
& 100.12 \\
\hline
\end{tabular}
}
\makebox[\textwidth][c]{%
\parbox{\textwidth}{%
\tablefoot{Results are reported for the full dataset (scenario~1) and for the dataset after subtracting the thermal component (scenario~2), as described in Sect.~\ref{ssection:agmodeling}. The first block lists parameters that are kept fixed during the fit. Posterior values correspond to the 16th, 50th, and 84th percentiles of the marginalized posterior distributions.}
}%
}
\end{table}

\begin{table}[]
    \caption{Emission line flux measurements of XRF~241001A host galaxy.}
    \centering
    \begin{tabular}{cc}
         \hline
         Emission line & Flux \\
         \hline
         \OII & $13.3 \pm 3.9$ \\
         \hbeta & $5.3 \pm 1.3$ \\
         \OIIIa & $13.4 \pm 1.3$ \\
         \OIIIb & $47.0 \pm 1.5$ \\
         \halpha & <$44.0$ \\
         \NIIb & <$38.0$ \\
         \hline
    \end{tabular}
    \label{tab:host_em_line_flux}
\tablefoot{Fluxes are given in units of $10^{-18}~\rm erg/s/cm^2$, are corrected for Galactic extinction ($A_V^{\rm Gal} = 0.03$ mag) and include a correction factor of 2.1 owing to the photometric calibration. Upper limits are given at the $3\sigma$ confidence level.}
\end{table}

\FloatBarrier

\end{document}